\documentclass[twocolumn]{aastex62}

\shorttitle{}
\shortauthors{Boer et al.}

\begin{document}

\title{Absence of a runaway greenhouse limit on lava planets}

\correspondingauthor{Iris D. Boer}
\email{irisboer07@gmail.com}

\author[0009-0004-4772-381X]{Iris D. Boer}
\affiliation{Kapteyn Astronomical Institute, University of Groningen, The Netherlands}

\author[0000-0002-8368-4641]{Harrison Nicholls}
\affiliation{Atmospheric, Oceanic and Planetary Physics, University of Oxford, United Kingdom}

\author[0000-0002-3286-7683]{Tim Lichtenberg}
\affiliation{Kapteyn Astronomical Institute, University of Groningen, The Netherlands}

\begin{abstract}

Climate transitions on exoplanets offer valuable insights into the atmospheric processes governing planetary habitability. Previous pure-steam atmospheric models show a thermal limit in outgoing long-wave radiation, which has been used to define the inner edge of the classical habitable zone and guide exoplanet surveys aiming to identify and characterize potentially habitable worlds. We expand upon previous modelling by treating (i) the dissolution of volatiles into a magma ocean underneath the atmosphere, (ii) a broader volatile range of the atmospheric composition including H$_2$O, CO$_2$, CO, H$_2$, CH$_4$ and N$_2$, and (iii) a surface temperature- and mantle redox-dependent equilibrium chemistry. We find that multi-component atmospheres of outgassed composition located above partially or fully-molten mantles do not exhibit the characteristic thermal radiation limit that arises from pure-steam models, thereby undermining the canonical concept of a runaway greenhouse limit, and hence challenging the conventional approach of using it to define an irradiation-based habitable zone. Our results show that atmospheric heat loss to space is strongly dependent on the oxidation and melting state of the underlying planetary mantle, through their significant influence on the atmosphere's equilibrium composition. This suggests an evolutionary hysteresis in climate scenarios: initially molten and cooling planets do not converge to the same climate regime as solidified planets that heat up by external irradiation. Steady-state models cannot recover evolutionary climate transitions, which instead require self-consistent models of the temporal evolution of the coupled feedback processes between interior and atmosphere over geologic time.

\end{abstract}


\section{Introduction} \label{sec:intro}
Exoplanets provide unique opportunities to explore the diverse phases and evolutionary trajectories of different planetary types throughout their lifetimes. Central to understanding the rich tapestry of planetary atmospheres is the study of atmospheric phenomena, such as the runaway greenhouse effect \citep{komabayasi-1967, ingersoll-1969, kasting-1988, Abe1988EvolutionOA, 1992JAtS...49.2256N}. Analyzing how the greenhouse effect manifests across different regimes provides us with critical insights into the factors influencing planetary habitability and evolutionary pathways \citep{2022ARA&A..60..159W, 2024arXiv240504057L}. A greater knowledge of the climatic evolution of exoplanets is necessary for interpreting current and future observations more accurately \citep{2024RvMG...90..411K}. Recent theoretical and observational studies have demonstrated that the atmospheric composition of low-mass exoplanets is tightly interconnected with the planetary interior and affects exoplanet observables \citep{2021ApJ...909L..22K,2021ApJ...922L...4D,2022PSJ.....3..127S,2024ApJ...962L...8S,2024arXiv240303325B,2024Natur.630..609H,2024NatCo..15.8374K}. The planetary interior can affect observational signatures of exoplanets in a number of ways; most important for the purpose of this manuscript is the degassing and/or ingassing of volatiles gases, which can influence the planetary transit radius and atmospheric composition \citep{2021ApJ...922L...4D,2021ApJ...914L...4L,2022PSJ.....3..127S}. If interaction with the interior can dominate the atmospheric composition, then it is natural to ask how this process may affect long-term climatic processes that are originally extrapolated from the study of Earth and the terrestrial planets of the Solar System. 

The `runaway greenhouse' (or `thermal') limit is the outgoing long-wave radiation (OLR) threshold that a planet with a pure-steam atmosphere is thought to approach for a particular range of surface temperatures. This limit can occur on initially temperate planets due to increased warming from their host stars \citep{Kopparapu_2013}. The blanketing effect of the atmosphere means that it absorbs thermal radiation from the surface and insulates the planet's interior such that the surface temperature can increase through a positive feedback loop as surface oceans are evaporated (in turn increasing the atmospheric opacity). This process is also important for setting the cooling of planets from an initially hot post-runaway state \citep{abematsui1985, kasting-1988}. The atmospheric temperature ($T$-$P$) profile is also intimately related to the phases of the volatiles throughout the atmosphere. Condensing gases may be rained-out \citep{pierrehumbert-2010}, which impacts the transport of radiation. Of greater importance, however, is that when the partial pressure of a given volatile is higher than its saturation vapour pressure, this volatile condenses and releases latent heat. Latent heat released during volatile condensation locally warms the atmosphere and decreases the steepness of the atmospheric lapse rate from a dry adiabat to a moist pseudioadiabat.
\par 
Under the often-invoked assumption of a purely convective temperature profile \citep{abematsui1985, 2013JGRE..118.1155L, boukrouche-2021, lichtenberg_2021}, an atmosphere dominated by a single volatile (e.g. steam) will greatly controlled by the latent heat release regions of condensation, which is entirely determined by the thermodynamic properties of that single volatile, and not by the surface temperature. This is crucial because when the photosphere of the atmosphere falls within this saturated region, the OLR cannot increase in step with the surface temperature \citep{pierrehumbert-2010,goldblatt2013,salvador2017relative, 2013Natur.497..607H}. This decoupling of radiating temperature from the surface temperature causes runaway warming of the planet. As the temperature further increases and the volatiles become supercritical, the dry adiabat reaches upwards to lower pressures such that the region of condensation becomes transparent. This restores the relationship between the upper atmosphere temperature and the surface, allowing for the OLR to increase with increasing surface temperature -- the so-called post-runaway stage \citep{boukrouche-2021}. This process -- the central behavior behind the runaway greenhouse limit -- can in principle operate for any volatile gas in an atmosphere that can be present in gaseous and condensed form at the surface, and is particularly critical for water vapour due to its large enthalpy of vaporisation, strong infrared opacity \citep{2010ppc..book.....P} and availability. 
In addition to the gas speciation itself, higher total atmospheric pressure increases atmospheric opacity, reducing OLR, while thinner atmospheres allow more radiation to escape into space \citep[e.g.,][]{boukrouche-2021}. This is called the runaway greenhouse effect and has classically been studied under terrestrial conditions, where water is the dominant condensable on the climatic conditions of Earth \citep[e.g.][]{komabayasi-1967, ingersoll-1969, kasting-1988, Abe1988EvolutionOA, 1992JAtS...49.2256N}.

\subsection{Implications of the runaway greenhouse limit}
Previous studies of the runaway greenhouse effect \citep[e.g.][]{Kopparapu_2013,goldblatt2013,Hamano_2015,boukrouche-2021,selsis-2023} modelled a terrestrial planet with an atmosphere composed purely out of steam. This pure-steam assumption stems from the pioneering work of Y. Abe and T. Matsui in the early 1980s \citep{abematsui1985}. 
These pure-steam models yield a runaway greenhouse threshold: a plateau representing approximately constant values of OLR for a range of surface temperatures. \cite{Kopparapu_2013} found a critical OLR value of approximately 293 W m$^{-2}$ for surface temperatures of 230 K to 1900 K. \cite{selsis-2023} found an OLR of approximately 276 W m$^{-2}$ for surface temperatures of 563 K to 1750 K. These represent the regime of the runaway greenhouse effect when a terrestrial planet is subject to increased irradiation; e.g. the future Earth as the Sun's luminosity increases with time. The onset point of this regime has previously been used to define the inner edge of the habitable zone \citep[e.g.][]{kasting1993habitable,Kopparapu_2013,zhangyang}, which is traditionally defined as a region around a star within which a planet with a CO$_2$-H$_2$O-N$_2$ atmosphere can maintain surface temperatures that allow for liquid water \citep{huang-1959,Hart1978,kasting1993habitable,UnderWoodJones2003,selsis2007b,KalteneggerSasselov2011}. The specific boundaries of the habitable zone depend on various factors, including the host star's luminosity, spectral type, the planet's mass, and the planet's atmospheric properties, which have been studied within the literature \citep[e.g.,][]{2011ApJ...734L..13P,2014ApJ...797L..25R,2017ApJ...837L...4R,2018ApJ...858...72R,2019ApJ...881..120K,2020MNRAS.494..259R,2021MNRAS.504.1029B,2022A&A...658A..40C,2023A&A...679A.126T}. Several of these works have performed studies on variations of the runaway greenhouse limit if the atmosphere is composed of a different background gas on the order of a few bar at maximum, and added specific gases in the H-C-N-O composition space. However, the majority of work in this area so far has arbitrarily varied the atmospheric conditions, mostly with specific settings in mind.

\subsection{Objective of this study}

With the perspective of using atmospheric models that help distinguish habitable worlds from non-habitable ones, here we alleviate some of the limitations of previous works. This focuses mainly on three interconnected aspects:

\begin{enumerate}
    \item Inclusion of a molten mantle and surface (a ``magma ocean'') which controls the atmospheric mass and composition through the dissolution of volatile species into the melt.
    \item Broadening the atmospheric compositional range by incorporating H$_2$O, CO$_2$, CO, H$_2$, CH$_4$, and N$_2$.
    \item Calculation of the equilibrium atmospheric composition and pressure based on the combined effects of equilibrium chemistry and dissolution into the melt (which are surface temperature dependent).
\end{enumerate}

A molten surface is considered for two key reasons. Firstly, planets that undergo a runaway greenhouse transition will have a dramatic increase in their surface temperature which is sufficient to melt the mantle \citep[e.g.,][]{2011E&PSL.304..251A}. Global radiative equilibrium within the post-runaway greenhouse state (when out- and incoming radiation balance each other) therefore will typically be accompanied by an underlying magma ocean \citep{2024arXiv240504057L}. Secondly, rocky planets are thought to start their lives with global magma oceans inherited from the accretionary heat of formation (for example by pebble accretion or giant impacts), which should thus be considered to be the starting condition for assessing climate evolution. The prevalence of the magma ocean phase is supported by comprehensive geological evidence \citep{tonks-1993,warren-1985,elkins-tanton-2008,nakajima-2021,2023RvMG...89...53C,2021ChEG...81l5735C,2023ASPC..534..907L} on Earth, the Moon, Mars \citep{2007Natur.450..525D}, and Venus \citep{2023SSRv..219...51S}, and is predicted both by planetesimal-based \citep[e.g.,][]{Alibert2013,2016ApJ...821..126Q, Wyatt2019,Emsenhuber2021,Schlecker2021,2020PSJ.....1...18C,2022ApJ...938L...3L} and pebble accretion models \citep[e.g.,][]{Morbidelli2012a,Lambrechts2019b,2021SciA....7..444J,2023A&A...671A..75J,2022E&PSL.58717537O} of planetary formation. Irrespective of the accretion path of rocky planets, strong irradiation by exoplanetary host stars \citep{2009A&A...506..287L,2021ChEG...81l5735C} or substantial tidal heating \citep{2015AsBio..15..739D,2020ApJ...900...24W,2025ApJ...979..133F} can cause rocky planetary mantles to melt for prolonged time periods, in addition to the additional greenhouse forcing of the atmosphere. Rapid energy loss on more temperate orbits could result in these magma ocean phases being relatively short, although different atmospheric compositions can induce a blanketing effect that extends the magma ocean phase for much longer \citep{lichtenberg-2021}.
The presence of a magma ocean adds an emerging feedback effect to the atmosphere's equilibrium composition, namely the solubility of gases into liquids, which allows for efficient volatile exchange between interior and atmosphere \citep[e.g.,][]{2016ApJ...829...63S,2018RSPTA.37680109S,bower2019linking,bower-2022,sossi-2023}, and critically links the amount of greenhouse gases in the atmosphere to the total amount of melt present in the deep interior \citep{2023FrEaS..1159412S,2024arXiv240504057L}. Considering the coupled feedback between magma oceans and their overlying atmosphere is therefore essential for understanding the climate and geological regimes of rocky planets, which cannot be captured by atmosphere-only models.
 
Incorporating more volatiles is necessary because pure-steam atmospheres are an extreme end-member case of oxidized planets -- such as the present-day Earth -- where low C/H ratios and high oxygen fugacities are the norm. The wide array of observed exoplanet systems makes it essential to model more diverse compositions applicable to a wider range of planetary systems. 

Lastly, considering surface temperature dependency is crucial for determining atmospheric equilibrium composition and the outgoing radiation. Equilibrium chemistry and the solubility of volatiles in the magma ocean are both determined based on temperature and pressure at the surface. Previous models considered mostly pure-steam compositions and studied deviations from this scenario, allowing for no interior-atmosphere volatile exchange (assuming planets with a completely solid surface). This likely does not fully capture the role of surface temperature on the equilibrium composition of the atmosphere and hence the OLR. However, for a multi-species atmosphere above a hot and molten surface, this temperature dependency is expected to play a more significant role of the established atmosphere due to complex equilibrium chemistry and interior-atmosphere volatile exchange. 

Therefore, we here study the dependency of the runaway greenhouse limit on atmospheric conditions that arise from interaction with the planetary interior. We aim to understand how the outgassed secondary atmosphere's composition, temperature, and pressure affect the OLR and hence the distinction between potentially habitable and thermally hot climate states \citep{2023SSRv..219...51S, 2024arXiv240504057L}. 

In Section \ref{sec:methods}, we describe the methods of our study. The results are presented in Section \ref{sec:results}, with additional data taken from the model to portray a broader picture of the observed concepts. We discuss their significance, implications and limitations in Section \ref{sec:discussion}; and conclude in Section \ref{sec:conclusions}.

\section{Methods} \label{sec:methods}

\subsection{Overview of methods}
In this study we combine equilibrium chemistry, volatile solubility, and atmospheric thermodynamics and radiative transfer calculations in order to examine the relation between the OLR and the oxidation state of the mantle. We model these planets with multi-species atmospheres located above a magma ocean. For the atmospheric temperature profiles, we stay as close as possible to the main assumptions applied in previous works which have studied the runaway greenhouse limit. Here we outline the procedure we follow within these models.

With the aim of extending upon the pure-steam atmospheric model, we make use of several components of \href{https://github.com/FormingWorlds/PROTEUS}{PROTEUS}: a coupled atmosphere-interior framework that models the evolution of rocky planets, which is derived from \citet{lichtenberg-2021} and has been significantly modified by \citet{harrison-2024, nicholls_convective_2024}. This framework includes the \href{https://github.com/FormingWorlds/JANUS}{JANUS} atmospheric module which is a 1-D prescribed atmosphere model of convective planetary atmospheres \citep{2021PSJ.....2..207G,2022JGRE..12707456G}, and the \href{https://github.com/FormingWorlds/CALLIOPE}{CALLIOPE} outgassing module, which is built on the methods of \citet{bower-2022,2024ApJ...962L...8S}. JANUS calculates atmospheric temperature and radiative fluxes according to surface volatile partial pressures determined by CALLIOPE.

\subsection{Equilibrium chemistry} 
\label{subsec:eqchem}
Assuming fixed H-C-N abundances and surface oxygen fugacity (which essentially defines the abundance of O atm), the atmosphere's surface composition and pressure can be calculated. Equilibrium chemistry includes the following redox reactions and their corresponding equilibrium constants $K_{\text{eq}}$ \citep[from JANAF and IVTANTHERMO,][]{chase-1998,2017ApJ...843..120S}:
\begin{equation}
    \text{H}_2 + 0.5\text{O}_2 \rightleftharpoons \text{H}_2\text{O},
    \label{H2O}
\end{equation}

\begin{equation}
    \quad \log K_{\text{eq}} = \frac{3.039 \times 10^4}{T} - 13152,
    \label{KH2O}
\end{equation}

\begin{equation}
    \text{CO} + 0.5\text{O}_2 \rightleftharpoons \text{CO}_2,
    \label{CO2}
\end{equation}

\begin{equation}
    \quad \log K_{\text{eq}} = \frac{4.348 \times 10^5}{T} - 14468,
    \label{KCO2}
\end{equation}

\begin{equation}
    \text{CO}_2 + 2\text{H}_2 \rightleftharpoons \text{CH}_4 + \text{O}_2,
    \label{CH4}
\end{equation}

\begin{equation}
    \quad \log K_{\text{eq}} = \frac{5.4738 \times 10^6}{T} - 16276.
    \label{K_CH4}
\end{equation}

Subject to these reactions, the dissolved volatile abundance is calculated using empirically derived solubility laws. The solubility constants for each volatile are taken from different solubility models, which are experimentally determined \citep{2023FrEaS..1159412S}. For H$_2$O we use the solubility model of \cite{sossi-2023}, which assumes a peridotite type of molten rock. Solubility models based on a basaltic type of molten rock are used for CO$_2$ \citep{dixon1995}, CO \citep{armstrong2015}, N$_2$ \citep{LIBOUREL20034123,2022GeCoA.336..291D} and CH$_4$ \citep{ARDIA201352}. The solubility of H$_2$ is not considered in this model since compared to the solubility constant of H$_2$O, that of H$_2$ is more than two orders of magnitude lower \citep{hirschmann2012solubility,Li2015}, making it negligible \citep{bower-2022}.

Newton's method is then applied to obtain the accurate corresponding atmospheric partial pressures that satisfy mass conservation of the input elemental abundances, equilibrium chemistry, and dissolution into the magma ocean. For a more detailed explanation of this, we refer to \citet{bower-2022}. With the obtained equilibrium composition conditions at the surface, the $T$-$P$ profile of the atmosphere is determined with JANUS utilising the multi-component non-dilute pseudoadiabat of \citet{2021PSJ.....2..207G}. We assume full rain-out of condensible species in the atmosphere.

\begin{figure}[htb!]
    \centering
    \includegraphics[width=0.99\linewidth]{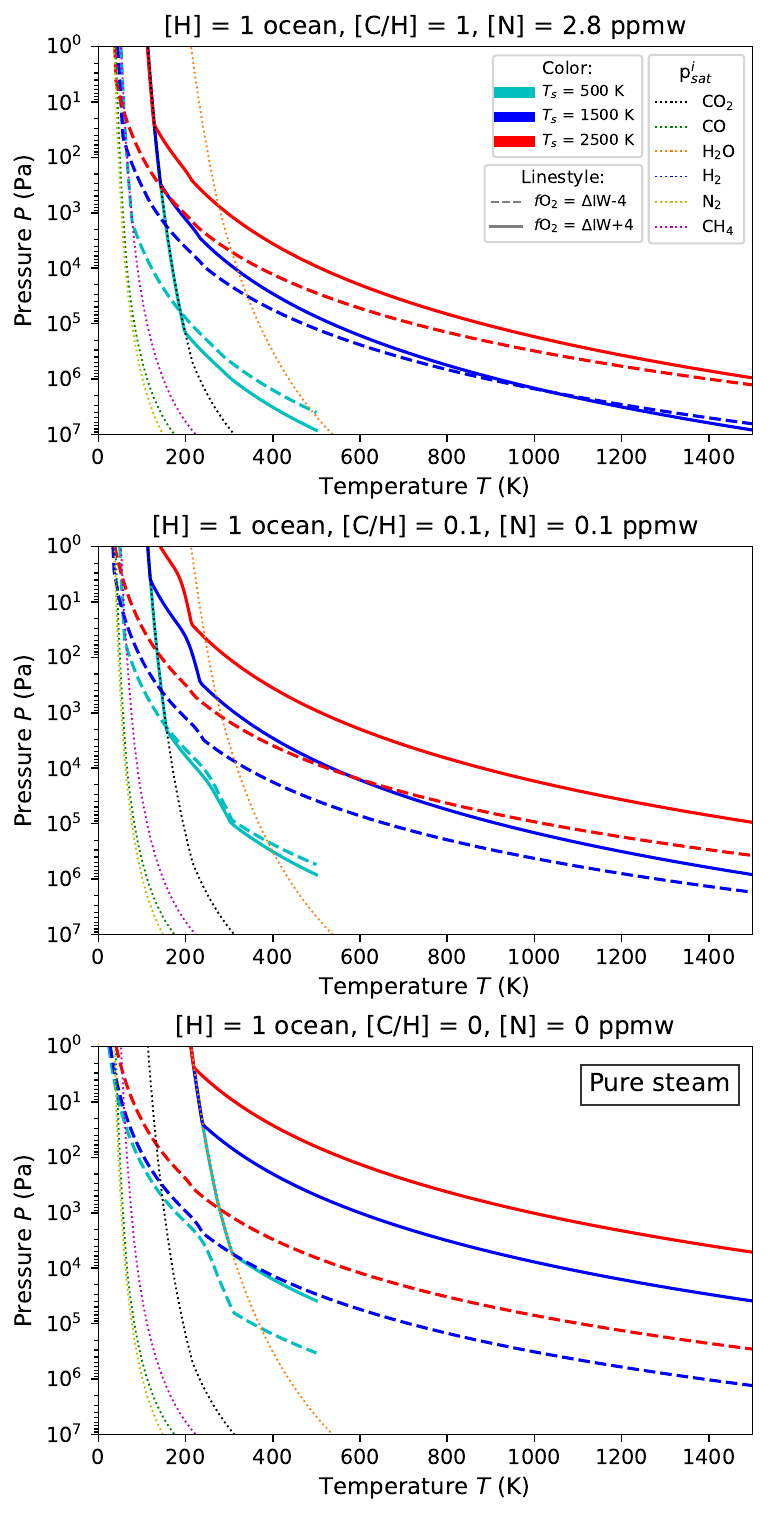}
    \caption{$T$-$P$ profiles showing the atmospheres established at various surface temperatures (line color) and redox states (line style). The steep line that all profiles follow in the upper atmospheres are due to the saturation vapour pressure curve. Saturation vapour pressure curves of different included volatiles are shown with dotted lines \citep{2010ppc..book.....P}. At deeper levels (higher pressures), the profiles follow the different dry adiabatic lapse rates which are minorly sensitive to atmospheric composition. These results are obtained with the \emph{open} degassing scenario of Figure \ref{fig:sketch}, where composition varies with temperature.}
    \label{fig:T-P}
\end{figure}

\subsection{Radiative transfer}
Given the atmospheric equilibrium composition and $T$-$P$ profile, radiative transfer calculations are applied to calculate the outgoing long-wave radiation (OLR). We use the JANUS module for this, which makes use of the atmospheric radiative transfer code `SOCRATES' \cite[][Suite Of Community Radiative Transfer codes based on Edwards and Slingo]{1996QJRMS.122..689E}. SOCRATES solves the radiative transfer equations via the two-stream, correlated-$k$, and plane-parallel approximations. A more extended description can be found in \citet{lichtenberg-2021}. Under these assumptions, the OLR is independent of the incoming stellar radiation. Opacity coefficients are taken from the HITRAN2020 database, making use of all available line and collision absorption coefficients for H$_2$O, H$_2$, CO$_2$, CO, CH$_4$ and N$_2$ \citep{2022JQSRT.27707949G,2023JQSRT.30608645M}. 

\begin{figure*}[htb!]
    \centering
    \includegraphics[width=1\linewidth]{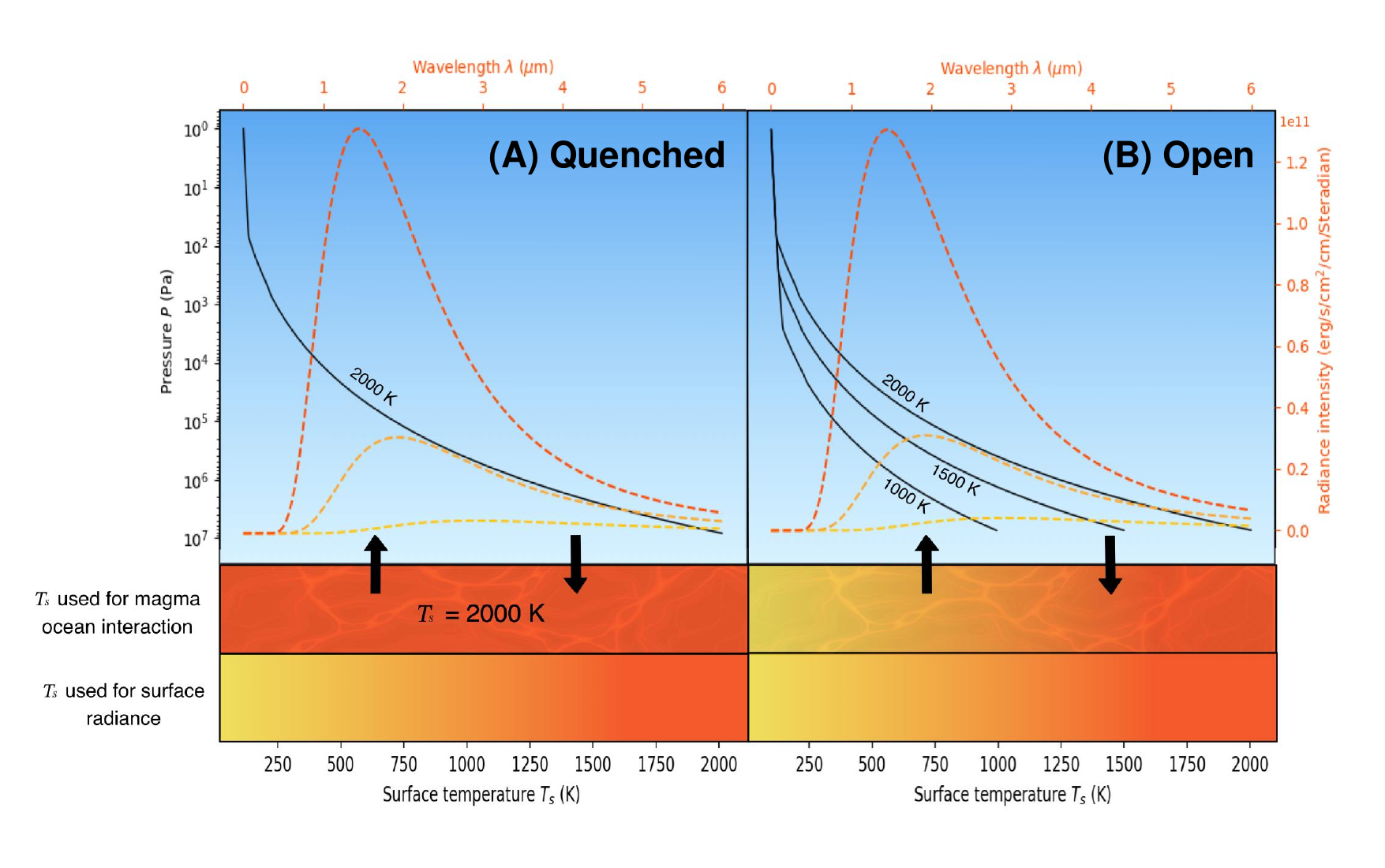}
    \caption{Illustration of the two different degassing scenarios used in this work, which bracket the potential pathways of mantle melting and crystallization of planetary evolution. The range of orange colours corresponds to temperatures as defined on the bottom abscissa. The figures contain $T$-$P$ profiles that are shown in black, with a surface temperature as defined by their lowermost endpoint; the corresponding black-body radiation curves are presented by striped lines in shades of orange and their radiance intensity and wavelength range are defined on the upper abscissa and right ordinate (in orange text). The black arrows denote the interaction between the atmosphere and the planetary mantle when determining the equilibrium composition of the atmosphere, as described in the Methods section. \textbf{(A)} The \emph{quenched} scenario is representative of a case where the primordial magma ocean equilibrates last with the atmosphere above the liquidus, i.e., when the entire mantle is molten. In this case, the chemical systems of the mantle and atmosphere are assumed to be 'closed' and no further in- or outgassing is taking place at lower surface temperatures. \textbf{(B)} The \emph{open} scenario represents the opposite end-member assumption: (partially) molten mantle and atmosphere are allowed to equilibrate at all surface temperatures, i.e., in- and outgassing is taking place over the whole temperature range, and the mantle is assumed to be an \emph{open} system towards the atmosphere.}
    \label{fig:sketch}
\end{figure*}

We neglect the effects of Rayleigh scattering, clouds/hazes, and the potential effects of a stratosphere, in order to enable a like-to-like comparison with central previous works. The lack of a stratosphere assumes that the atmosphere is convective throughout and the temperature decreases monotonically with height \cite{nicholls_convective_2024}. We discuss the impact of these assumptions in Section \ref{sec:limitations}.

\subsection{Modelling scenarios} 
For a given input of H-, C- and N-abundance, redox state, and melt fraction of the mantle, we determine the OLR, equilibrium composition and $T$-$P$ profile of the corresponding atmosphere that is established in equilibrium above a magma ocean. This model is used to obtain the data presented in Section \ref{sec:results} and archived on Zenodo\footnote{\href{https://doi.org/10.5281/zenodo.14002553}{https://doi.org/10.5281/zenodo.14002553}} for reproduction purposes. 

Throughout this paper, the redox state is consistently expressed on a decadic logarithmic scale relative to the iron-w\"ustite buffer, which is the equilibrium reaction between iron (Fe) and w\"ustite (FeO) at given $P$-$T$ conditions \citep[see, e.g.,][]{2023ASPC..534..907L,2023SSRv..219...51S}. Henceforth, we will denote
log$_{10}$($f$O$_2$/IW) as $\Delta$IW$\pm$X for simplicity. To study the redox dependency of the OLR, the oxygen fugacity (a proxy for mantle redox state) is varied over a range of $\Delta$IW+4 (indicating an oxidized state close to the Earth's lithosphere at present day) to $\Delta$IW-4 (indicating a reduced state, closer to core-forming conditions or Mercury's surface at the present day).
Throughout this paper, we simulate various atmospheres by varying parameters that represent the H-abundance in terms of Earth's oceans (in mole), the C-abundance in terms of C/H ratio, and the N-abundance in ppmw relative to the mantle mass. Default values of these parameters are often used, when emphasis is placed on one specific parameter, default values of the other parameters are set to a default value. These default values are based on Earth: [H] = 1 ocean, [C/H] = 1, [N] = 2.8 ppmw and $f$O$_2$ = $\Delta$IW+4.

The influence of the surface temperature and the redox state on the atmospheric thermal structure is shown in Figure \ref{fig:T-P}: $T$-$P$ profiles are presented corresponding to different mantle redox states and surface temperatures. Different combinations of redox state (through the oxygen fugacity) and surface temperature result in different temperature profiles, implying different radiative behaviour through the location of the photosphere and the saturated region of the atmosphere, as described in Section \ref{sec:intro}. The redox-dependency of these $T$-$P$ profiles is set by the redox-dependency of the equilibrium composition, which will be discussed later in Section \ref{sec:results}. 

To study the significance of the surface temperature dependency of the equilibrium chemistry, we outline two different scenarios. For each run of set parameters of H-, C- and N-abundance and the redox state, we determine two different equilibrium compositions using these two scenarios, and calculate the OLR for both. The two different scenarios are the \emph{quenched} and \emph{open} degassing scenarios, which are illustrated in Figure \ref{fig:sketch}. The difference between these two scenarios is the surface temperature used to determine the composition of the atmosphere. The \emph{quenched} scenario (Figure \ref{fig:sketch}, left) involves modeling an atmospheric composition which is independent of changes in the surface temperature, similar to the approach adopted in previous models; in this scenario we use a fixed temperature of 2000 K, as seen in Figure \ref{fig:sketch}. However, the OLRs for the quenched scenarios are still calculated for the whole range of surface temperatures, but using this specific quenched atmospheric composition. In these cases, the surface temperatures impacts the OLR through the atmospheric temperature profile's control over the radiative fluxes. The surface radiance is determined with a black-body radiation curve. For the \emph{open} scenario (Figure \ref{fig:sketch}, right), we set the model to re-determine the composition of the atmosphere that allows the system to be self-consistently in equilibrium with each surface temperature value. Consequently, for each surface temperature, the OLR is calculated based on an atmosphere composition consistent with the surface temperature through the equilibrium chemistry of the volatiles (see Section \ref{subsec:eqcompo}). Therefore, for the OLR calculation using the \emph{open} scenario, the utilized equilibrium gas composition is dependent on the change in surface temperature. We discuss the implications and limitations of this approach in sections \ref{sec:discussion} and \ref{sec:limitations}.

\begin{figure*}[htb!]
    \centering
    \includegraphics[width=1\linewidth]{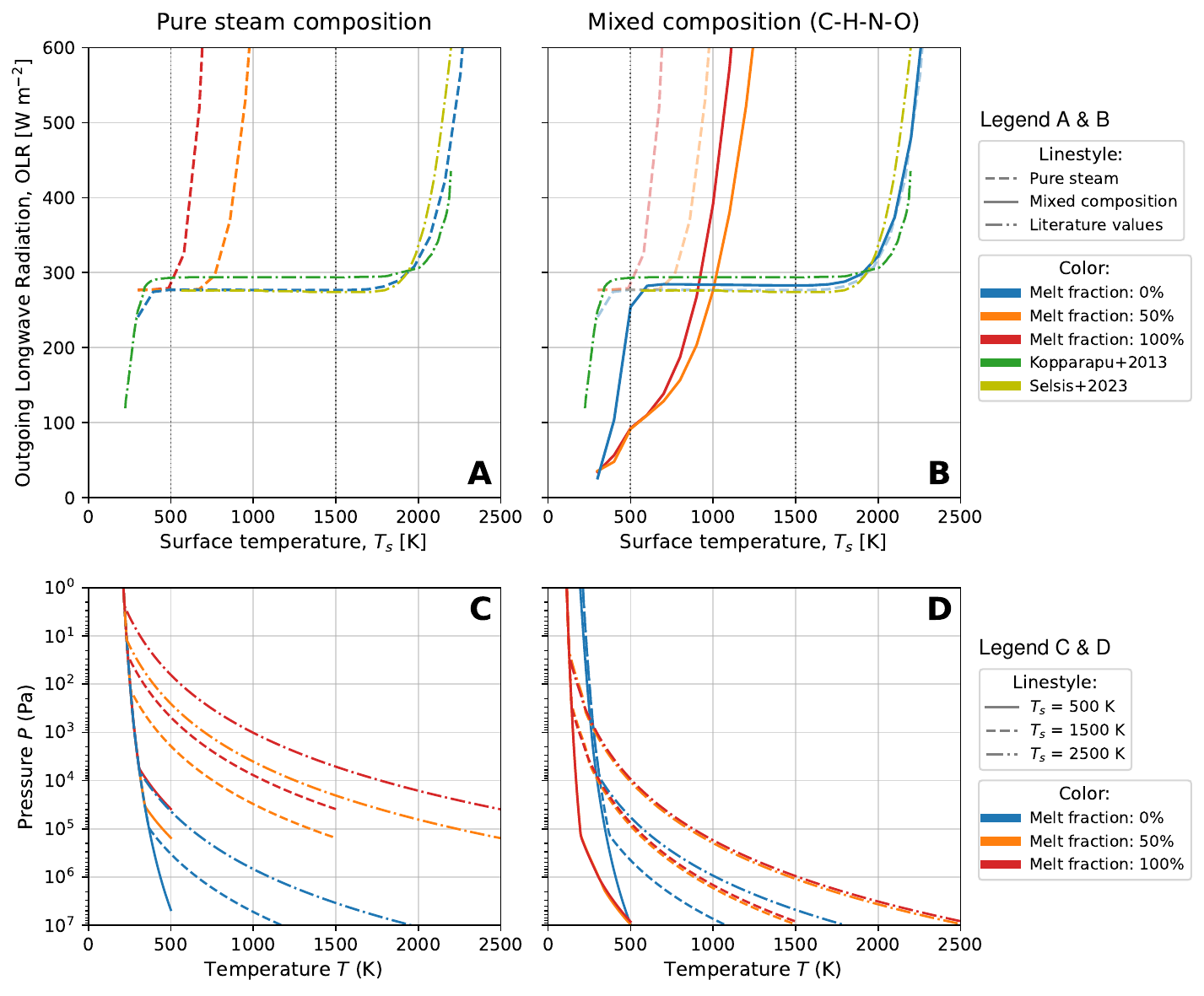}
    \caption{$T$-$P$ profiles showing the atmospheres established at various surface temperatures above a planetary interior with different melt fractions (line color).
    Outgoing long-wave radiation (OLR) for various surface temperatures, melt fractions, and chemical equilibrium compositions. \textbf{(A)} Pure-steam atmospheres above planetary interiors with different melt fractions of 0\% (blue), 50\% (orange) and 100\% (red), with [H] = 1 Earth ocean, [C/H] = 0, [N] = 0 ppmw of the mantle mass, $f$O$_2$ = $\Delta$IW+4. Pure-steam cases from the models of \cite{Kopparapu_2013} (green dash-dotted line) and \cite{selsis-2023} (yellow dash-dotted line) are shown for comparison with a solid surface case (blue dashed line). \textbf{(B)} Cases with chemically equilibrated compositions in the H-C-N system above planetary mantles with different melt fractions, with [H] = 1 ocean, [C/H] = 1, [N] = 2.8 ppmw, $f$O$_2$ = $\Delta$IW+4. Models from (A) are shown for comparison purposes as faint dashed lines. Note that the blue dashed and solid lines almost completely overlap. \textbf{(C)} $T$-$P$ profiles of the three pure-steam atmospheres with different planetary interior melt fractions (line color) shown in (A) at surface temperatures of 500 K, 1500 K and 2500 K (line type). \textbf{(D)} $T$-$P$ profiles of the three mixed-composition atmospheres with different planetary interior melt fractions (line color) shown in (B) at surface temperatures of 500 K, 1500 K and 2500 K (line type).}
    \label{fig:MO}
\end{figure*}
\section{Results} \label{sec:results}

First, we show how the OLR behaviour changes when deviating from the previous pure-steam assumption. Secondly, we describe the redox dependency of the equilibrium atmospheric composition. Finally, we highlight the OLR behaviour of different climate regimes across redox states, element inventory and their surface temperature dependency. All figures include the OLR curve from \cite{Kopparapu_2013} to indicate the differences introduced by our modelling with JANUS with respect to the previously studied pure-steam cases.

\subsection{Influence of mantle melt fraction on thermal limit}

We first test the JANUS model under the pure-steam assumption, with results compared to literature values from \cite{Kopparapu_2013} and the adiabatic case from \cite{selsis-2023}. Figure \ref{fig:MO}A shows the dashed blue line for JANUS, aligned closely with other models from 300 K to 2000 K, where a constant OLR can be observed across this range: the key characteristic of the runaway greenhouse limit. Beyond 2000 K the models diverge slightly, with the OLR calculated by \cite{selsis-2023} increasing beyond 1950 K, while JANUS and \cite{Kopparapu_2013} increase beyond 2000 K. At 2450 K, JANUS further deviates, with an OLR difference of up to 100 W m$^{-2}$ from \cite{selsis-2023}. 

Looking further into the pure-steam case (Figure \ref{fig:MO}A), one can see that the relation between the OLR and the surface temperature changes significantly when the surface underneath the atmosphere is (partly) molten (see dashed orange and red lines). In these cases the OLR increases beyond the runaway limit at much lower temperatures: approximately 500 K for the fully molten case and 750 K for the 50\% melt fraction case. The variation in melt fraction yields different surface pressures because the molten interior serves as a dissolution sink for volatiles. This is shown in Figure \ref{fig:MO}C, where planetary interiors with a higher melt fraction consistently result in atmospheric structures with lower (surface) pressures (see solid orange and red lines). The three displayed melt fraction cases in Figure \ref{fig:MO}A) have thus different surface pressures: at 500 K the 0\% melt fraction yields 269 bar, 50\% yields 1.5 bar, and 100\% gives 0.4 bar. For all three cases, the surface pressure shows a minimal variation with temperature (for a given melt fraction): from 500 K to 3000 K, the maximum decrease of surface pressure is $\sim$1.5\%. This decrease in total surface pressure for an increasing melt fraction is caused by H being increasingly partitioned into the mantle by the solubility law and therefore being less present in the atmosphere, which results in a substantially changed atmospheric structure and hence OLR. These low pressures for atmospheres overlying the molten interiors do not allow for condensation in the atmosphere and are inefficient at absorbing radiation emitted from the surface of the planet, thereby generating much larger OLR values for a given surface temperature, compared to the 0\% melt fraction case.

When C and N are included in the planet's volatile inventory, other volatiles beside H$_2$O are formed in the atmosphere: CO$_2$, CO, H$_2$, CH$_4$ and N$_2$. The results from these more mixed-composition models (2.8 ppmw of N, C/H = 1, $\Delta$IW+4) are shown in Figure \ref{fig:MO}B. The pure-steam results from are shown as dashed lines in this panel for comparison, highlighting how the addition of further compounds affects the OLR of molten cases. At 500 K, a solid interior results in a surface pressure of 240 bar with an H$_2$O-dominated atmosphere; at 50\% and 100\% melt fractions, the surface pressure drops to 97 bar and 84 bar respectively, with CO$_2$ being the dominant gas. Unlike in the pure-steam case, atmospheric surface pressures vary with surface temperature, reaching 376 bar for the 0\% melt fraction, 95 bar for 0\% melt fraction, and 82.7 bar for 100\% melt fraction at 2000 K. 
The molten surface conditions in case (B) that include a mixed atmospheric composition do not exhibit an OLR plateau at all. However, the OLR-temperature relationship for solid surfaces remains unchanged between the (A) and (B) cases, except below 500 K, where the (B) case shifts the OLR to smaller values between 300 and 500 K, deviating from the solid pure-steam case. This shift occurs because, below 500 K, the mixed composition atmosphere has a cooler temperature profile than the pure steam atmosphere at similar pressures (as shown in Figure \ref{fig:T-P}). As the pure steam atmosphere follows the saturation vapour pressure curve of H$_2$O aloft, it remains hotter, resulting in higher OLR values.

\begin{figure*}[htb!]
    \centering
    \includegraphics[width=0.99\linewidth]{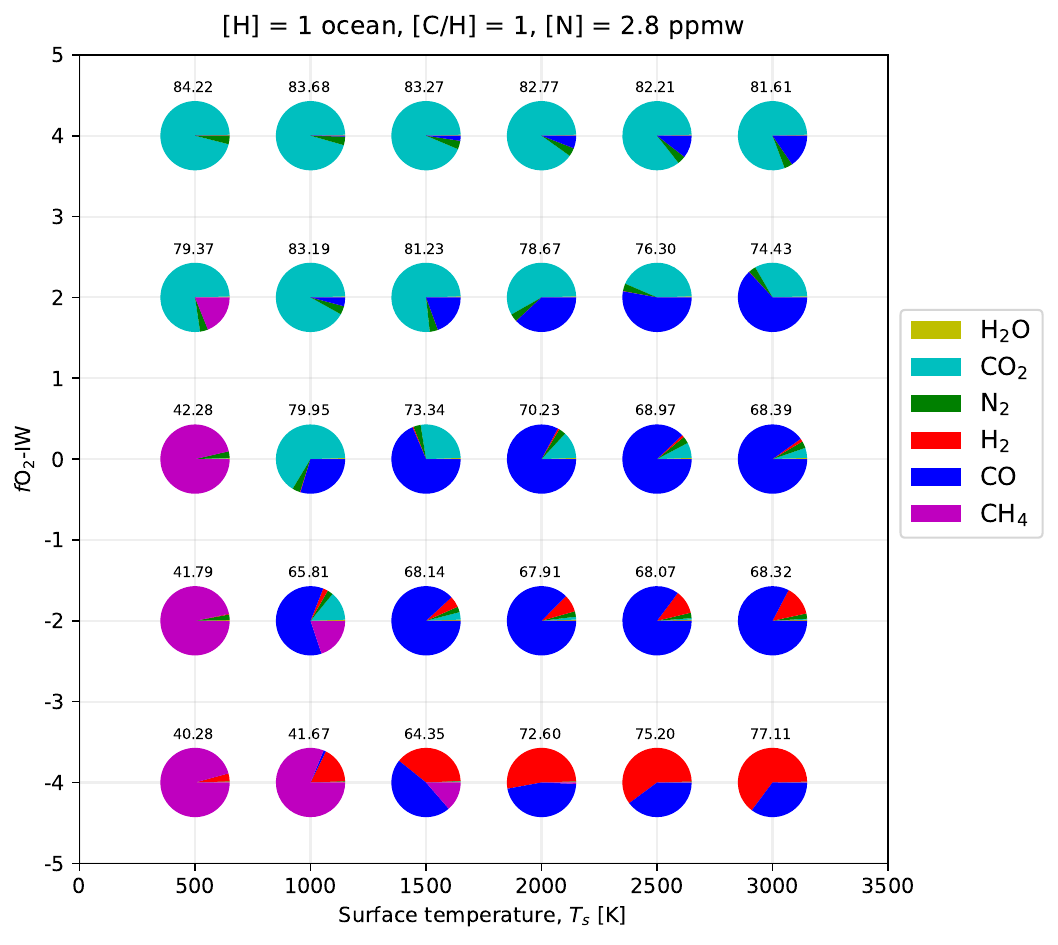}
    \caption{Atmospheric composition (volume mixing ratio) above a magma ocean for different redox states and surface temperatures. The total pressure (in bar) of each atmosphere is given above the corresponding pie chart. Other parameters are kept at the default parameter values and the \emph{open} degassing scenario from Figure \ref{fig:sketch} is used.}
    \label{fig:piechart}
\end{figure*}

\subsection{Equilibrium atmospheric composition degassed from a magma ocean}
\label{subsec:eqcompo}

A significant parameter in the formation of the atmosphere is the redox state of the planetary interior. When the mantle is in an oxidizing state, more O-rich molecules (here, H$_2$O and CO$_2$) will form than O-poor molecules (N$_2$, H$_2$, CO and CH$_4$). The atmospheric composition primarily determines the radiative behaviour of a planet, hence we first study how the redox-dependency affects the equilibrium composition. The sensitivity of the atmospheric composition to the redox state and surface temperature (through the equilibrium chemistry as explained in \ref{subsec:eqchem}) of a fully molten planetary mantle is shown in Figure \ref{fig:piechart}. The most significant feature of this composition map is the variety in dominant species across the range of redox states and surface temperatures, where, due to the strong solubility of H, none of the modeled atmospheres are dominated by H$_2$O, a feature of magma ocean-atmosphere equilibration that was pointed out before by \citet{bower-2022,2020SciA....6.1387S,sossi-2023}.
Atmospheres with similar dominant species exhibit surface pressures within comparable ranges. The dependency of the atmospheric composition on the equilibrium chemistry leads to a wide range of atmospheric compositions. Under oxidizing conditions ($\Delta$IW+0 to $\Delta$IW+4), CO$_2$-dominated atmospheres are prevalent across a broad range of temperatures, with total surface pressures typically between 79 and 84 bar. For more reducing conditions between $\Delta$IW-2 and $\Delta$IW+0, outgassed atmospheres with surface temperatures above 1000 K are generally CO-dominated, with total surface pressures ranging from 65 to 74 bar. Moderately oxidizing outgassed atmospheres ($\Delta$IW+2) at high temperatures may also be CO-dominated above 2000 K. CH$_4$-dominated atmospheres occur in reducing conditions ($\Delta$IW-4 to $\Delta$IW+0) at 500 and 100 K, with total surface pressures $\sim 41 \text{ bar}$ (ranging from 40.28 to 42.28 bar). H$_2$-dominated atmospheres appear at $\Delta$IW-4 and temperatures above 2000 K; these atmospheres are primarily composed of CO and H$_2$, with total surface pressure negatively correlated with the mixing ratio of CO (due to its higher molar mass than H$_2$). Overall, H$_2$O is nearly absent in all atmospheres, and N$_2$ appears only in oxidizing atmospheres in small quantities. N$_2$ is hardly present within the most reduced atmospheres we have considered ($\Delta$IW-4) due to the high solubility of nitrogen species in magma at reducing conditions \citep{2024ApJ...962L...8S}; for more oxidizing atmospheres, N$_2$ remains only in small quantities due to the input volatile budget of N being relatively low compared to the input volatile budgets of C and H \citep{WANG2018460}.

\begin{figure*}[htb!]
    \centering
    \includegraphics[width=0.99\linewidth]{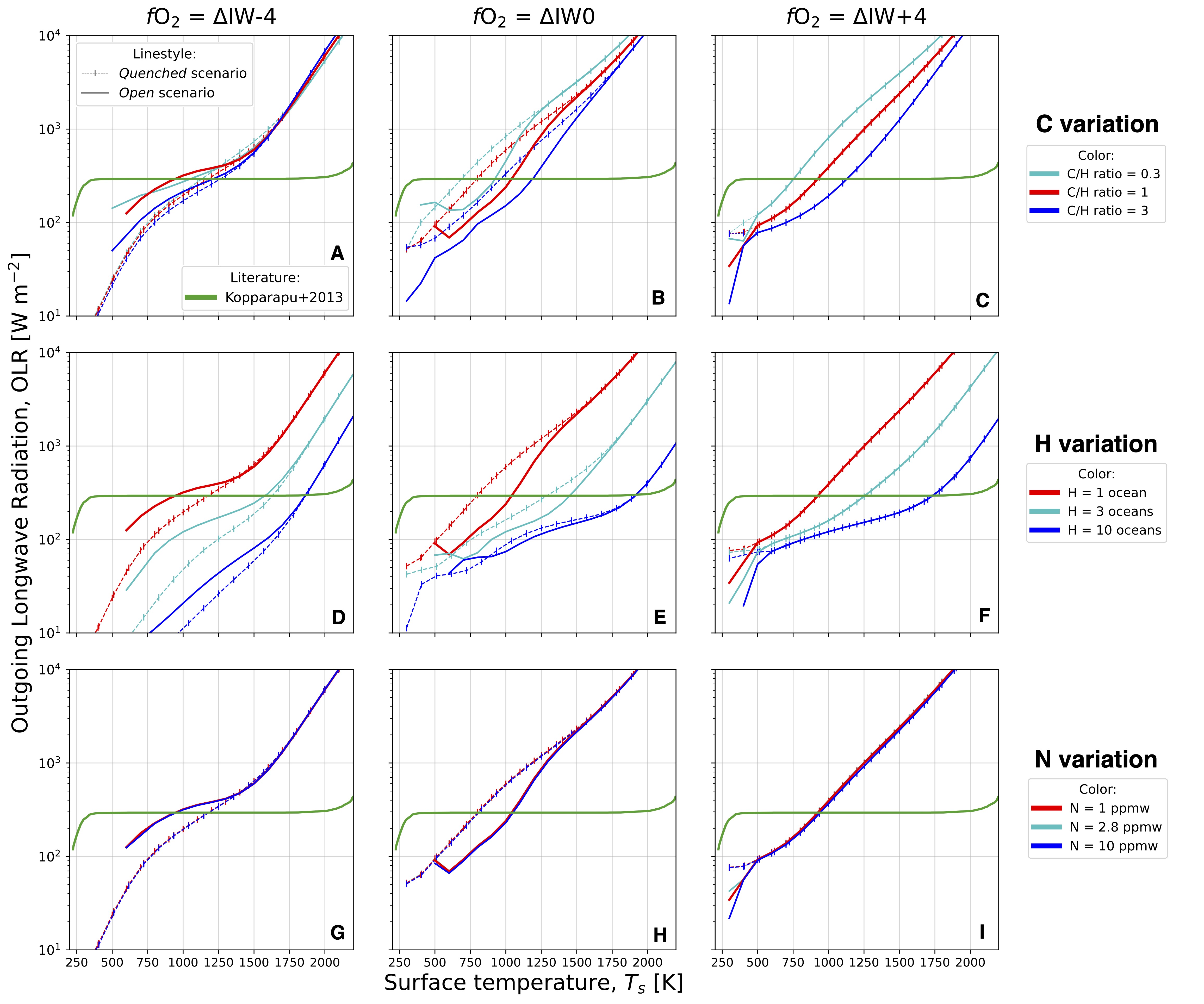}
    \caption{OLR vs. surface temperature for varied atmospheric equilibrium compositions. Graphs left to right show variations on C/H ratio (first row, A/B/C), H budget (second row, D/E/F) and N budget (third row, G/H/I) for different redox states ($\Delta$IW). Colours denote the volatile budgets and line types represent the used equilibrium composition scenarios. Untouched parameters are set to their default values}
    \label{fig:var9}
\end{figure*}

\subsection{Vanishing thermal limit from magma ocean-atmosphere coupling}
\label{subsec:OLR}

The sensitivity of atmospheric composition on the planetary oxidation state has a significant effect on the possible climate state of rocky planets, as shown in Figure \ref{fig:var9}. In this figure, we analyse the impact of outgassing temperature on the OLR using the \emph{quenched} (dashed lines) and \emph{open} (solid lines) degassing scenarios from Figure \ref{fig:sketch}. 

Focusing first on the \emph{open} scenario (solid lines), all the OLR curves in Figure \ref{fig:var9} show an increase in OLR with increasing surface temperature, although none of the cases exhibit an OLR plateau. All panels show OLR curves that start at a lower OLR value relative to the \cite{Kopparapu_2013} model, but surpass the pure-steam thermal limit at a surface temperatures ranging from approximately 900 K to 1750 K, then rapidly increasing above the thermal limit. Various panels show an inflection point after which the OLR curves increase their increasing rate, resulting into the OLR curves showing different behaviour across the surface temperature. 

With varying elemental abundances of H, C, and N, different atmosphere compositions show different characteristic radiative behaviour. In the case of C-abundance variations (Figures \ref{fig:var9} A, B, C) we see that higher C-to-H ratios lead to lower OLR values across all redox states. At $\Delta$IW-4, higher C/H ratios show first a relative shallow increase in OLR, followed by a rapid increase in OLR above $\approx$ 1500 K. For $\Delta$IW+0 and $\Delta$IW+4, the rate of increase is more gradual.

H-abundance variations (Figures \ref{fig:var9}D, E, F) show that larger amounts of hydrogen result in lower OLR values across all redox conditions. At $\Delta$IW-4, OLR exhibits an inflection point where it begins to increase rapidly. This inflection shifts to higher temperatures as H-abundance increases. At $\Delta$IW+0 and $\Delta$IW+4, a similar pattern is observed, but the onset of rapid OLR increase occurs at different temperatures depending on the H-abundance.

Lastly, when varying N-abundance (Figures \ref{fig:var9}G, H, I) the effect on OLR is less pronounced compared to H and C. At $\Delta$IW-4, a similar initial shallow increase of OLR is seen, but overall, OLR increases steadily with temperature across all redox conditions. Variations in N-abundance cause only minor deviations in OLR, particularly at lower temperatures, while the effect is minimal at higher temperatures.  

Considering the differences between the \emph{quenched} and \emph{open} scenarios, the surface temperature dependency of the OLR becomes more clear. The \emph{quenched} OLR curves (dashed) in Figure \ref{fig:var9} for $\Delta$IW-4 generally follow the same pattern as the \emph{open} curves (solid) at lower temperatures, but at a reduced OLR level, with a maximum deviation of 75 W/m$^2$ at 600 K. The curves converge around 1500 K and remain nearly identical at higher temperatures. In general, for certain atmospheres, the \emph{quenched} curves show lower OLR at lower temperatures and converge with the \emph{open} curves around 1500 K. However, for higher H-abundances convergence occurs at 1750 K. The deviation in OLR between the \emph{quenched} and \emph{open} curves is more pronounced for variations in H-abundances (Figures \ref{fig:var9}D, E, F) than for C/H ratios or N-abundances (Figures \ref{fig:var9}A, B, C, G, H and I).

For $\Delta$IW+0, \emph{quenched} curves show generally higher OLR levels compared to the \emph{open} curves, with inflection points at around 1500 K across all figures. Again, H-abundance variations display greater deviations between the two scenarios compared to C/H and N-abundance cases.

\section{Discussion} \label{sec:discussion}
Figure \ref{fig:MO} validates our JANUS atmosphere model by the comparison of its results with previous work, demonstrating that it reproduces the pure-steam OLR limit. The minor deviations from the \citet{Kopparapu_2013} and \citet{selsis-2023} models likely stem from JANUS using surface temperature-dependent equilibrium composition (\emph{open} scenario), while the reference models use a constant surface pressure of 270 bar. Additionally, the models use different linelists and heat capacities. Despite these differences, the alignment of the OLR curves demonstrates sufficient accuracy and reliability for the use cases in this study.

\subsection{The absence of a thermal radiation limit}

\subsubsection{Influence of a magma ocean}
The addition of a magma ocean underneath the atmosphere adds additional physics and chemistry to the model system. Figure \ref{fig:MO} shows that the dissolution of volatiles in the magma ocean can be significant, and the melt fraction of the mantle can thereby strongly influence the relation between the OLR and the surface temperature. 

The effects of changing the mantle melt fraction are shown in Figure \ref{fig:MO} and can be explained by the resultant differences in atmospheric composition and $T$-$P$ profile. Dissolution of volatiles into the underlying magma ocean significantly changes the surface pressure. The pure-steam case with a solid mantle has an average surface pressure of 269 bar, whereas a fully molten mantle obtains an average surface pressure of 0.4 bar; representing an extremely thin atmosphere. The absorption of radiation by the atmosphere influences how much radiation will escape to space. This explains why the atmospheres located above a magma ocean radiate significantly more energy to space at these lower temperatures compared to the models without volatile dissolution into a magma ocean.

The OLR curves do not change linearly with the melt fraction. This is related to the non-linear behaviour of the temperature profiles of the atmosphere. A larger melt fraction allows for more volatiles to dissolve in the mantle, hence lowering the partial pressure of that volatile and therefore the total pressure in the atmosphere. However, this does not translate directly to the upper atmosphere -- where the photosphere is approximately located -- because of the non-linear shape $T(p)$ of the pseudoadiabatic temperature profiles.

The atmospheres of mixed composition (Figure \ref{fig:var9} do not first attain super-runaway OLR fluxes at the same surface temperatures. Lower melt fractions (corresponding to larger surface pressures) transition into a post-runaway state at higher temperatures because the region of moist convection is much deeper, which requires higher temperatures for the dry adiabat to span to the radiating level. Given that a terrestrial planet starts with a magma ocean (and a large melt fraction), this implies that it could spend more time in a post-runaway state than previously thought (compare the red and blue curves in Figure \ref{fig:MO}, panel A). These higher long-wave fluxes could result in a shorter cooling timescale, although evolutionary calculations are required to explore this further.

\subsubsection{Mixed atmospheric compositions}
Comparing the pure-steam (panel A) and mixed composition (panel B) graphs in Figure \ref{fig:MO}, we can see that the pure-steam atmosphere is relatively more sensitive to the mantle melt fraction. This is because its atmosphere consists wholly of H$_2$O, which is significantly more soluble in magma than the other volatiles modelled in this work \citep{2023FrEaS..1159412S}. Since the mixed-composition atmospheres feature a relatively lower abundance of H$_2$O, these scenarios are less affected by melt fraction variations. For the pure-steam cases, this strong sensitivity to dissolution into the interior results in a stronger decrease in opacity with respect to the atmospheres with a mixed composition. Incorporating a more diverse set of volatiles introduces more complexity in contrast to the pure steam atmosphere. The addition of other volatiles increases the sensitivity of atmospheric conditions (surface pressure and composition) to surface temperature and melt fraction, leading to large changes in OLR, as shown in Figure \ref{fig:MO}. The strong infrared opacity and relatively high dew point temperature of steam -- both of which are fundamental to generating the radiation plateau of the canonical runaway greenhouse -- are both undermined by the introduction of a partially molten surface (which readily dissolves the H$_2$O) and the addition of other volatiles into the gas mixture (which do not have the same radiative and thermodynamic properties as H$_2$O). These differences are central to understanding why the thermal radiation limit is not present for atmospheres of mixed composition.

The impact of H-, C-, and N-abundance variations on OLR varies in magnitude. Increasing H-abundance lowers OLR for a given surface temperature due to higher atmospheric opacity -- largely through the collisional absorption of H$_2$ \citep{harrison-2024} --  as reflected in the rightward shift of the OLR curve in Figure \ref{fig:var9} D, E and F. C-abundance variations (Figure \ref{fig:var9} A, B and C) show less shift, partly because the C/H ratio is based on a fixed H-abundance of 1 Earth ocean, whereas H-abundance -- which is varied independently -- span a wider range of 1 to 10 oceans. Additionally, the preferential formation of CO or CO$_2$ results in more similar equilibrium compositions across C-abundances, decreasing variation in radiative behaviour.

The studied range of N variations, given in ppmw relative to the mantle mass, has no significant impact on the OLR due to relatively smaller variation over concentrations and the absence of other nitrogen-bearing species like NH$_3$ that have some infrared opacity. In contrast to N, our model does include oxidising H and C species, which substantially add to the atmospheric opacity. Several factors contribute to the opacity, including the additional transitions that are associated with the oxygen bonds in H$_2$O and CO$_2$. Additionally, an increase in the pressure leads to larger opacities due to collisional absorption and pressure-broadening of the lines. This factor additionally contributes to the result showing lower OLR values for the higher element inventories of H and C. Finally, the collision-induced absorption of CO is unknown \citep{2022JQSRT.27707949G}, which thus represents an empirical bottleneck for studying magma ocean climates at reduced conditions.

\subsubsection{Temperature dependency of the atmospheric composition}

As discussed in Section \ref{sec:results}, the OLR is dependent on the surface temperature used for equilibrium chemistry, especially under reduced conditions at temperatures below 1500 K. The differences between the \emph{quenched} and \emph{open} scenarios at these temperatures can be attributed to the composition of the atmosphere. Lower temperature cases allow for deeper regions of moist convection in the atmosphere and associated rain-out of volatiles. Once a volatile condenses within a given layer of the atmosphere, cold-trapping means that it has a volume mixing ratio of zero at that level and above, meaning that it does not contribute to the radiative transfer in those regions. 

This is consistent with the results in Figure \ref{fig:T-P}, showing deep regions of moist convection at higher pressures for atmospheres established at lower surface temperatures. Results presented under the \emph{open} scenario differ from those of the \emph{quenched} quenched scenario, because the former scenario calculates the gas composition self-consistently with the surface temperature, whereas the  latter fixes the composition to that corresponding to 2000 K. This results in different OLR curves between the two scenarios, all else equal, and thereby highlights the importance of a temperature-dependent calculation of volatile speciation from an underlying magma ocean.

The same principle applies to redox variations. The redox state influences the equilibrium atmospheric composition (see Figure \ref{fig:piechart}), controls the specific temperature-pressure profile (see Figure \ref{fig:T-P}) and simultaneously the phase (i.e. condensed or vapour) of the volatiles, and consequently the OLR. The redox state of the underlying planetary mantle is therefore also a significant factor in determining an atmosphere's radiative behaviour, as previously shown by \citet{harrison-2024}.

There are smaller differences in the OLR fluxes between the \emph{open} and \emph{quenched} scenarios at higher temperatures, which suggests that the composition of the atmosphere has less impact on the OLR in this regime. This is because at higher surface temperatures the atmospheric temperature profiles generally follow a dry adiabat -- as can be seen in Figure \ref{fig:T-P} -- meaning that condensation is not occurring, hence the radiative behaviour of the atmosphere is less affected by any difference between the compositions based on the two different scenarios. The dry adiabatic lapse rate is minorly controlled by the composition of the atmosphere -- through the difference in specific heat capacity and mean molecular weight -- but this is a second-order effect compared to the role of moist convection \citep{selsis-2023, pierrehumbert-2010}.

In general, reduced atmospheres ($\Delta$IW-4) yield a lower OLR using the \emph{quenched} scenario with respect to the \emph{open} scenario, and oxidized atmospheres ($\Delta$IW+4) result in similar OLR for both scenarios. This pattern is explained by Figure \ref{fig:piechart}, which shows how equilibrium compositions change with surface temperature and redox state. For $\Delta$IW+4, the atmosphere remains CO$_2$-dominated at all temperatures, resulting in similar OLR for both cases. In contrast, at $\Delta$IW+0, the composition shifts from CO$_2$ to CO and CH$_4$ dominance as temperature decreases, affecting radiative behaviour. This is a result of the temperature dependence of the carbon redox reactions modelled in this work.

\subsection{Implications}
The runaway greenhouse threshold that arises from a pure-steam atmosphere is absent when an underlying magma ocean and redox-dependent atmospheric composition are modelled self-consistently. With the absence of this thermal limit, the inner edge of the classical habitable zone can no longer be defined based solely on irradiation from the host star: the surface habitability of rocky planets also strongly depends on the composition of the atmosphere set by the volatile endowment of the planet, and the phase and redox state of the underlying mantle. Determining surface liquid water conditions requires detailed knowledge of the atmosphere's composition and temperature profile, and cannot be tied simply to the instellation that a planet receives. This suggests that initially molten planets cooling down from a magma ocean episode after planetary formation cannot be divided into irradiation-based orbital zones \citep[e.g.,][]{2021JGRE..12606643K,2023A&A...679A.126T}. Previous climate models (including GCMs) similarly have found varying influences on the thermal radiation limit due to various processes, such as clouds, additional backgrounds gases, and variable rotation rate  \citep{2011ApJ...734L..13P,2013ApJ...771L..45Y,2014ApJ...797L..25R,2017ApJ...837L...4R,2018ApJ...858...72R,2019ApJ...881..120K,2020MNRAS.494..259R,2021MNRAS.504.1029B,2021Natur.598..276T,2022A&A...658A..40C,selsis-2023,2023A&A...679A.126T}.

Critically, however, none of these studies so far has included the implications of a dominantly molten surface and redox-dependent chemistry on the outgoing steady-state radiation. Rocky planets form in highly energetic states due to the release of gravitational potential energy, giant impacts, tidal heating, and radiogenic heating during planetary formation \citep{2023ASPC..534..907L,2023ASPC..534.1031K}. Therefore, whether planets form a potentially habitable surface and climate is determined by if and how they transition from a hot to a colder and more temperate state. Our simulations including molten mantles exhibit no trend of a radiation limit, and hence a simplistic classification by irradiation must be assessed in the full compositional and energetic parameter space possible for rocky planets. Given the strong redox dependence of greenhouse limits, future studies should explore the effects of H, C, and N abundances, oxygen fugacity, and stellar radiation on resultant atmospheric equilibrium composition and radiation environment. 

The dependence of OLR on the composition and melt state of a planet will likely impact the cooling rate of its mantle. When a thermal limit is present, the OLR is independent of surface temperature, allowing planets to cool down as soon as the incoming stellar radiation is low enough. However, in our models, OLR varies with temperature, which can either speed up or slow down magma ocean cooling with respect to the magma ocean cooling rate of an atmosphere with a thermal limit. For instance, hydrogen-rich atmospheres with lower OLR values could prolong the magma ocean phase, while oxidized, carbon-poor atmospheres may shorten it. In recent years, several studies have started to explore a wider compositional range of magma ocean and climatic evolution \citep{lichtenberg-2021,bower-2022,2024NatCo..15.8374K}. The results presented in here underline that models considering only the climatic effects of a fixed atmospheric composition cannot assess the interdependent feedback between the planetary interiors and atmospheres. 

Tighter synchronization with exoplanet surveys is thus required to test the complex feedback between exoplanet interiors and atmospheres, and enable conclusions on planetary surface conditions based on astronomical observables. For example, previous models have suggested that steam runaway greenhouse states should lead to significant inflation of the atmospheric scale height \citep{2020A&A...638A..41T}, which can be tested with the near-future PLATO survey \citep{2024PSJ.....5....3S}. However, our results here predict that such a relationship should not exist, as interior dissolution and variable redox states will affect the equilibrium climate state of the variety of rocky exoplanets by orders of magnitude, effectively erasing this signature. In order to test our understanding of the early climatic evolution of rocky exoplanets, observations of the early transient magma ocean epoch of young planets are thus critical \citep{2019A&A...621A.125B,2024arXiv241013457C}.

\subsection{Limitations}
\label{sec:limitations}

Our study uses steady-state simulations to model the interaction of a planet's climate and its interior, making a number of simplifying assumptions. First, in order to enable a like-to-like comparison to previous climate models that were used to establish the runaway greenhouse hypothesis, we assume the atmosphere to be fully convective, following the multi-species moist pseudoadiabat derived by \citet{2021PSJ.....2..207G}, and ignore the potential formation of a stratosphere, Rayleigh scattering, and clouds. Clouds can either inhibit or enhance outgoing radiation by increasing opacity or reflecting incoming radiation \citep{marcq-2017}, whereas Rayleigh scattering will also contribute to the outgoing radiation by scattering incoming stellar radiation back to space. However, with less radiation reaching the surface, this also results in a smaller temperature increase at the surface and therefore less radiation coming from the surface that will be transferred towards the top of the atmosphere. Ultimately, these assumptions are limiting the applicability of our models to the wide phase space of exoplanets, however, for canonical habitable zone-like irradiations and orbits these are assumptions comparable to other works in this area. \citet{selsis-2023} challenged the assumption of a convective lower atmosphere, arguing that the equilibrium state of a steam atmosphere should lead to a solidified surface. Our study here cannot assess the impact of radiative zones and thus cannot be directly compared to these results. However, \citet{selsis-2023} ignore the dissolution of volatiles into the magma ocean at early times, hence their atmospheric conditions are arbitrarily chosen. Future work shall evaluate the intimate relation between magma ocean crystallization, degassing, atmospheric convection, and radiation. 

Finally, our main results, which model the dissolution of volatiles, assume a mantle melt fraction of 100\%, which is a simplification that impacts our results at temperatures below 1200 K to 1400 K, when basaltic rocks are expected to crystallize. However, in petrologically motivated models of magma ocean crystallization, the surface temperature and melt fraction of the mantle decouple \citep{2007Natur.450..866L,2017GGG....18.3385B,lichtenberg-2021,bower-2022,2023MNRAS.526.6235M}, and hence a mapping between surface temperature and melt fraction is not straightforward. Future work is required to establish self-consistent models to test against experimental results and exoplanet observations.

\section{Conclusions}
\label{sec:conclusions}
Previous models based of pure-steam atmospheres (or slight variations thereof) have traditionally defined the inner edge of the habitable zone by analyzing the thermal runaway limit of outgoing long-wave radiation. Our study extends upon these models by: (i) incorporating the presence of a possible magma ocean beneath the atmosphere, which allows the volatiles to dissolve and relates the atmospheric composition based on mantle oxidation state, (ii) expanding the range of atmospheric compositions considered to include H$_2$O, CO$_2$, CO, H$_2$, CH$_4$, and N$_2$, and (iii) a accounting for temperature dependent atmospheric chemistry. 

Our study reveals a substantial revision in the relationship between OLR and surface temperature compared to pure-steam and atmosphere-only models. Specifically, we observe that atmospheres with multi-species compositions situated above magma oceans do not exhibit a runaway greenhouse threshold at all. This is in contrast to the previously considered pure-steam cases and atmosphere-only models that have previously perturbed the composition relative to this canonical steam case. This insight strongly challenges the conventional notion of using stellar irradiation and OLR only to define orbital regions where liquid water would be possible on the surface of exoplanets.
Moreover, our results emphasize the strong redox sensitivity of atmospheric equilibrium composition, in particular due to the dissolution of volatiles into the planetary interior. 

Our findings thus suggest that initially molten planets evolve toward qualitatively different climate regimes than non-molten planets, suggesting a hysteresis effect in planetary evolution that requires time-evolved models to bridge exoplanet observations with planetary geophysics.

\acknowledgments
T.L. was supported by the Branco Weiss Foundation, the Netherlands eScience Center (PROTEUS project, NLESC.OEC.2023.017), the Alfred P. Sloan Foundation (AEThER project, G202114194), and NASA's Nexus for Exoplanet System Science research coordination network (Alien Earths project, 80NSSC21K0593). 

\software{ \href{https://github.com/FormingWorlds/PROTEUS}{PROTEUS} \citep{lichtenberg-2021,harrison-2024}, \href{https://github.com/FormingWorlds/JANUS/releases/tag/v24.04.04}{JANUS} \citep{2021PSJ.....2..207G,2022JGRE..12707456G}, \href{https://github.com/FormingWorlds/CALLIOPE}{CALLIOPE} \citep[based on][]{bower-2022,sossi-2023,2024ApJ...962L...8S}, \href{https://github.com/nichollsh/SOCRATES}{SOCRATES} \citep{1996QJRMS.122..689E}, {Numpy} \citep{harris2020array}, {Matplotlib} \citep{4160265}.
        }

\vspace{0.5cm}
\newpage
\bibliography{references}{}

\begin{thebibliography}{}
\expandafter\ifx\csname natexlab\endcsname\relax\def\natexlab#1{#1}\fi
\providecommand{\url}[1]{\href{#1}{#1}}
\providecommand{\dodoi}[1]{doi:~\href{http://doi.org/#1}{\nolinkurl{#1}}}
\providecommand{\doeprint}[1]{\href{http://ascl.net/#1}{\nolinkurl{http://ascl.net/#1}}}
\providecommand{\doarXiv}[1]{\href{https://arxiv.org/abs/#1}{\nolinkurl{https://arxiv.org/abs/#1}}}

\bibitem[{Abe \& Matsui(1985)}]{abematsui1985}
Abe, Y., \& Matsui, T. 1985, Journal of Geophysical Research: Solid Earth, 90,
  C545

\bibitem[{{Abe} \& {Matsui}(1988)}]{Abe1988EvolutionOA}
{Abe}, Y., \& {Matsui}, T. 1988, Journal of the Atmospheric Sciences, 45, 3081,
  \dodoi{10.1175/1520-0469(1988)045<3081:EOAIGH>2.0.CO;2}

\bibitem[{Alibert {et~al.}(2013)Alibert, Carron, Fortier, Pfyffer, Benz,
  Mordasini, \& Swoboda}]{Alibert2013}
Alibert, Y., Carron, F., Fortier, A., {et~al.} 2013, \aap, 558, 1,
  \dodoi{10.1051/0004-6361/201321690}

\bibitem[{{Andrault} {et~al.}(2011){Andrault}, {Bolfan-Casanova}, {Nigro},
  {Bouhifd}, {Garbarino}, \& {Mezouar}}]{2011E&PSL.304..251A}
{Andrault}, D., {Bolfan-Casanova}, N., {Nigro}, G.~L., {et~al.} 2011, EPSL,
  304, 251, \dodoi{10.1016/j.epsl.2011.02.006}

\bibitem[{Ardia {et~al.}(2013)Ardia, Hirschmann, Withers, \&
  Stanley}]{ARDIA201352}
Ardia, P., Hirschmann, M., Withers, A., \& Stanley, B. 2013, Geochimica et
  Cosmochimica Acta, 114, 52, \dodoi{https://doi.org/10.1016/j.gca.2013.03.028}

\bibitem[{Armstrong {et~al.}(2015)Armstrong, Hirschmann, Stanley, Falksen, \&
  Jacobsen}]{armstrong2015}
Armstrong, L.~S., Hirschmann, M.~M., Stanley, B.~D., Falksen, E.~G., \&
  Jacobsen, S.~D. 2015, Geochimica et Cosmochimica Acta, 171, 283

\bibitem[{{Benneke} {et~al.}(2024){Benneke}, {Roy}, {Coulombe}, {Radica},
  {Piaulet}, {Ahrer}, {Pierrehumbert}, {Krissansen-Totton}, {Schlichting},
  {Hu}, {Yang}, {Christie}, {Thorngren}, {Young}, {Pelletier}, {Knutson},
  {Miguel}, {Evans-Soma}, {Dorn}, {Gagnebin}, {Fortney}, {Komacek},
  {MacDonald}, {Raul}, {Cloutier}, {Acuna}, {Lafreni{\`e}re}, {Cadieux},
  {Doyon}, {Welbanks}, \& {Allart}}]{2024arXiv240303325B}
{Benneke}, B., {Roy}, P.-A., {Coulombe}, L.-P., {et~al.} 2024, arXiv e-prints,
  arXiv:2403.03325, \dodoi{10.48550/arXiv.2403.03325}

\bibitem[{{Bonati} {et~al.}(2019){Bonati}, {Lichtenberg}, {Bower}, {Timpe}, \&
  {Quanz}}]{2019A&A...621A.125B}
{Bonati}, I., {Lichtenberg}, T., {Bower}, D.~J., {Timpe}, M.~L., \& {Quanz},
  S.~P. 2019, \aap, 621, A125, \dodoi{10.1051/0004-6361/201833158}

\bibitem[{{Bonati} \& {Ramirez}(2021)}]{2021MNRAS.504.1029B}
{Bonati}, I., \& {Ramirez}, R.~M. 2021, \mnras, 504, 1029,
  \dodoi{10.1093/mnras/stab891}

\bibitem[{{Boukar{\'e}} \& {Ricard}(2017)}]{2017GGG....18.3385B}
{Boukar{\'e}}, C.~E., \& {Ricard}, Y. 2017, Geochemistry, Geophysics,
  Geosystems, 18, 3385, \dodoi{10.1002/2017GC006902}

\bibitem[{Boukrouche {et~al.}(2021)Boukrouche, Lichtenberg, \&
  Pierrehumbert}]{boukrouche-2021}
Boukrouche, R., Lichtenberg, T., \& Pierrehumbert, R.~T. 2021, The
  Astrophysical journal, 919, 130, \dodoi{10.3847/1538-4357/ac1345}

\bibitem[{Bower {et~al.}(2022)Bower, Hakim, Sossi, \& Sanan}]{bower-2022}
Bower, D.~J., Hakim, K., Sossi, P.~A., \& Sanan, P. 2022, The Planetary Science
  Journal, 3, 93, \dodoi{10.3847/psj/ac5fb1}

\bibitem[{Bower {et~al.}(2019)Bower, Kitzmann, Wolf, Sanan, Dorn, \&
  Oza}]{bower2019linking}
Bower, D.~J., Kitzmann, D., Wolf, A.~S., {et~al.} 2019, Astronomy \&
  Astrophysics, 631, A103

\bibitem[{{Canup} {et~al.}(2023){Canup}, {Righter}, {Dauphas}, {Pahlevan},
  {{\'C}uk}, {Lock}, {Stewart}, {Salmon}, {Rufu}, {Nakajima}, \&
  {Magna}}]{2023RvMG...89...53C}
{Canup}, R.~M., {Righter}, K., {Dauphas}, N., {et~al.} 2023, Reviews in
  Mineralogy and Geochemistry, 89, 53, \dodoi{10.2138/rmg.2023.89.02}

\bibitem[{{Cesario} {et~al.}(2024){Cesario}, {Lichtenberg}, {Alei},
  {Carri{\'o}n-Gonz{\'a}lez}, {Dannert}, {Defr{\`e}re}, {Ertel}, {Fortier},
  {Garc{\'\i}a Mu{\~n}oz}, {Glauser}, {Hansen}, {Helled}, {Huber}, {Ireland},
  {Kammerer}, {Laugier}, {Lillo-Box}, {Menti}, {Meyer}, {Noack}, {Quanz},
  {Quirrenbach}, {Rugheimer}, {van der Tak}, {Wang}, {Anger},
  {Balsalobre-Ruza}, {Bhattarai}, {Braam}, {Castro-Gonz{\'a}lez}, {Cockell},
  {Constantinou}, {Cugno}, {Davoult}, {G{\"u}del}, {Hernitschek}, {Hinkley},
  {Itoh}, {Janson}, {Johansen}, {Jones}, {Kane}, {van Kempen}, {Kislyakova},
  {Korth}, {Kovacevic}, {Kraus}, {Kuiper}, {Mathew}, {Matsuo}, {Miguel}, {Min},
  {Navarro}, {Ramirez}, {Rauer}, {Vow Ricketti}, {Romagnolo}, {Schlecker},
  {Sneed}, {Squicciarini}, {Stassun}, {Tamura}, {Viudez-Moreiras},
  {Wordsworth}, \& {the LIFE Collaboration}}]{2024arXiv241013457C}
{Cesario}, L., {Lichtenberg}, T., {Alei}, E., {et~al.} 2024, arXiv e-prints,
  arXiv:2410.13457, \dodoi{10.48550/arXiv.2410.13457}

\bibitem[{{Chao} {et~al.}(2021){Chao}, {deGraffenried}, {Lach}, {Nelson},
  {Truax}, \& {Gaidos}}]{2021ChEG...81l5735C}
{Chao}, K.-H., {deGraffenried}, R., {Lach}, M., {et~al.} 2021, Chemie der Erde
  / Geochemistry, 81, 125735, \dodoi{10.1016/j.chemer.2020.125735}

\bibitem[{Chase(1998)}]{chase-1998}
Chase, M. 1998, American Institute of Physics, -1

\bibitem[{{Chaverot} {et~al.}(2022){Chaverot}, {Turbet}, {Bolmont}, \&
  {Leconte}}]{2022A&A...658A..40C}
{Chaverot}, G., {Turbet}, M., {Bolmont}, E., \& {Leconte}, J. 2022, \aap, 658,
  A40, \dodoi{10.1051/0004-6361/202142286}

\bibitem[{{Clement} {et~al.}(2020){Clement}, {Kaib}, \&
  {Chambers}}]{2020PSJ.....1...18C}
{Clement}, M.~S., {Kaib}, N.~A., \& {Chambers}, J.~E. 2020, \psj, 1, 18,
  \dodoi{10.3847/PSJ/ab91aa}

\bibitem[{{Dasgupta} {et~al.}(2022){Dasgupta}, {Falksen}, {Pal}, \&
  {Sun}}]{2022GeCoA.336..291D}
{Dasgupta}, R., {Falksen}, E., {Pal}, A., \& {Sun}, C. 2022, \gca, 336, 291,
  \dodoi{10.1016/j.gca.2022.09.012}

\bibitem[{{Debaille} {et~al.}(2007){Debaille}, {Brandon}, {Yin}, \&
  {Jacobsen}}]{2007Natur.450..525D}
{Debaille}, V., {Brandon}, A.~D., {Yin}, Q.~Z., \& {Jacobsen}, B. 2007, \nat,
  450, 525, \dodoi{10.1038/nature06317}

\bibitem[{Dixon \& Pan(1995)}]{dixon1995}
Dixon, J.~E., \& Pan, V. 1995, American Mineralogist, 80, 1339,
  \dodoi{10.2138/am-1995-11-1224}

\bibitem[{{Dorn} \& {Lichtenberg}(2021)}]{2021ApJ...922L...4D}
{Dorn}, C., \& {Lichtenberg}, T. 2021, \apjl, 922, L4,
  \dodoi{10.3847/2041-8213/ac33af}

\bibitem[{{Driscoll} \& {Barnes}(2015)}]{2015AsBio..15..739D}
{Driscoll}, P.~E., \& {Barnes}, R. 2015, Astrobiology, 15, 739,
  \dodoi{10.1089/ast.2015.1325}

\bibitem[{{Edwards} \& {Slingo}(1996)}]{1996QJRMS.122..689E}
{Edwards}, J.~M., \& {Slingo}, A. 1996, Quarterly Journal of the Royal
  Meteorological Society, 122, 689, \dodoi{10.1002/qj.49712253107}

\bibitem[{Elkins-Tanton(2008)}]{elkins-tanton-2008}
Elkins-Tanton, L. 2008, Earth and Planetary Science Letters, 271, 181,
  \dodoi{10.1016/j.epsl.2008.03.062}

\bibitem[{Emsenhuber {et~al.}(2021)Emsenhuber, Mordasini, Burn, Alibert, Benz,
  \& Asphaug}]{Emsenhuber2021}
Emsenhuber, A., Mordasini, C., Burn, R., {et~al.} 2021, \aap, 656, A69,
  \dodoi{10.1051/0004-6361/202038553}

\bibitem[{{Farhat} {et~al.}(2025){Farhat}, {Auclair-Desrotour}, {Bou{\'e}},
  {Lichtenberg}, \& {Laskar}}]{2025ApJ...979..133F}
{Farhat}, M., {Auclair-Desrotour}, P., {Bou{\'e}}, G., {Lichtenberg}, T., \&
  {Laskar}, J. 2025, \apj, 979, 133, \dodoi{10.3847/1538-4357/ad9b93}

\bibitem[{{Goldblatt} {et~al.}(2013){Goldblatt}, {Robinson}, {Zahnle}, \&
  {Crisp}}]{goldblatt2013}
{Goldblatt}, C., {Robinson}, T.~D., {Zahnle}, K.~J., \& {Crisp}, D. 2013,
  Nature Geoscience, 6, 661, \dodoi{10.1038/ngeo1892}

\bibitem[{{Gordon} {et~al.}(2022){Gordon}, {Rothman}, {Hargreaves}, {Hashemi},
  {Karlovets}, {Skinner}, {Conway}, {Hill}, {Kochanov}, {Tan}, {Wcis{\l}o},
  {Finenko}, {Nelson}, {Bernath}, {Birk}, {Boudon}, {Campargue}, {Chance},
  {Coustenis}, {Drouin}, {Flaud}, {Gamache}, {Hodges}, {Jacquemart}, {Mlawer},
  {Nikitin}, {Perevalov}, {Rotger}, {Tennyson}, {Toon}, {Tran}, {Tyuterev},
  {Adkins}, {Baker}, {Barbe}, {Can{\`e}}, {Cs{\'a}sz{\'a}r}, {Dudaryonok},
  {Egorov}, {Fleisher}, {Fleurbaey}, {Foltynowicz}, {Furtenbacher}, {Harrison},
  {Hartmann}, {Horneman}, {Huang}, {Karman}, {Karns}, {Kassi}, {Kleiner},
  {Kofman}, {Kwabia-Tchana}, {Lavrentieva}, {Lee}, {Long}, {Lukashevskaya},
  {Lyulin}, {Makhnev}, {Matt}, {Massie}, {Melosso}, {Mikhailenko}, {Mondelain},
  {M{\"u}ller}, {Naumenko}, {Perrin}, {Polyansky}, {Raddaoui}, {Raston},
  {Reed}, {Rey}, {Richard}, {T{\'o}bi{\'a}s}, {Sadiek}, {Schwenke},
  {Starikova}, {Sung}, {Tamassia}, {Tashkun}, {Vander Auwera}, {Vasilenko},
  {Vigasin}, {Villanueva}, {Vispoel}, {Wagner}, {Yachmenev}, \&
  {Yurchenko}}]{2022JQSRT.27707949G}
{Gordon}, I.~E., {Rothman}, L.~S., {Hargreaves}, R.~J., {et~al.} 2022, \jqsrt,
  277, 107949, \dodoi{10.1016/j.jqsrt.2021.107949}

\bibitem[{{Graham} {et~al.}(2021){Graham}, {Lichtenberg}, {Boukrouche}, \&
  {Pierrehumbert}}]{2021PSJ.....2..207G}
{Graham}, R.~J., {Lichtenberg}, T., {Boukrouche}, R., \& {Pierrehumbert}, R.~T.
  2021, \psj, 2, 207, \dodoi{10.3847/PSJ/ac214c}

\bibitem[{{Graham} {et~al.}(2022){Graham}, {Lichtenberg}, \&
  {Pierrehumbert}}]{2022JGRE..12707456G}
{Graham}, R.~J., {Lichtenberg}, T., \& {Pierrehumbert}, R.~T. 2022, J Geophys
  Res Planets, 127, e2022JE007456, \dodoi{10.1029/2022JE007456}

\bibitem[{{Hamano} {et~al.}(2013){Hamano}, {Abe}, \&
  {Genda}}]{2013Natur.497..607H}
{Hamano}, K., {Abe}, Y., \& {Genda}, H. 2013, \nat, 497, 607,
  \dodoi{10.1038/nature12163}

\bibitem[{Hamano {et~al.}(2015)Hamano, Kawahara, Abe, Onishi, \&
  Hashimoto}]{Hamano_2015}
Hamano, K., Kawahara, H., Abe, Y., Onishi, M., \& Hashimoto, G.~L. 2015, The
  Astrophysical Journal, 806, 216, \dodoi{10.1088/0004-637X/806/2/216}

\bibitem[{Harris {et~al.}(2020)Harris, Millman, Van Der~Walt, Gommers,
  Virtanen, Cournapeau, Wieser, Taylor, Berg, Smith,
  {et~al.}}]{harris2020array}
Harris, C.~R., Millman, K.~J., Van Der~Walt, S.~J., {et~al.} 2020, Nature, 585,
  357

\bibitem[{{Hart}(1978)}]{Hart1978}
{Hart}, M.~H. 1978, Icarus, 33, 23, \dodoi{10.1016/0019-1035(78)90021-0}

\bibitem[{Hirschmann {et~al.}(2012)Hirschmann, Withers, Ardia, \&
  Foley}]{hirschmann2012solubility}
Hirschmann, M.~M., Withers, A., Ardia, P., \& Foley, N. 2012, Earth and
  Planetary Science Letters, 345, 38

\bibitem[{{Hu} {et~al.}(2024){Hu}, {Bello-Arufe}, {Zhang}, {Paragas},
  {Zilinskas}, {van Buchem}, {Bess}, {Patel}, {Ito}, {Damiano}, {Scheucher},
  {Oza}, {Knutson}, {Miguel}, {Dragomir}, {Brandeker}, \&
  {Demory}}]{2024Natur.630..609H}
{Hu}, R., {Bello-Arufe}, A., {Zhang}, M., {et~al.} 2024, \nat, 630, 609,
  \dodoi{10.1038/s41586-024-07432-x}

\bibitem[{Huang(1959)}]{huang-1959}
Huang, S. 1959, Astronomical Society of the Pacific, 71, pp. 421.
\newblock \url{https://www.jstor.org/stable/40673575}

\bibitem[{Hunter(2007)}]{4160265}
Hunter, J.~D. 2007, Computing in Science \& Engineering, 9, 90,
  \dodoi{10.1109/MCSE.2007.55}

\bibitem[{{Ingersoll}(1969)}]{ingersoll-1969}
{Ingersoll}, A.~P. 1969, Journal of the Atmospheric Sciences, 26, 1191,
  \dodoi{10.1175/1520-0469(1969)026<1191:TRGAHO>2.0.CO;2}

\bibitem[{{Johansen} {et~al.}(2021){Johansen}, {Ronnet}, {Bizzarro},
  {Schiller}, {Lambrechts}, {Nordlund}, \& {Lammer}}]{2021SciA....7..444J}
{Johansen}, A., {Ronnet}, T., {Bizzarro}, M., {et~al.} 2021, Sci Adv, 7,
  eabc0444, \dodoi{10.1126/sciadv.abc0444}

\bibitem[{{Johansen} {et~al.}(2023){Johansen}, {Ronnet}, {Schiller}, {Deng}, \&
  {Bizzarro}}]{2023A&A...671A..75J}
{Johansen}, A., {Ronnet}, T., {Schiller}, M., {Deng}, Z., \& {Bizzarro}, M.
  2023, \aap, 671, A75, \dodoi{10.1051/0004-6361/202142142}

\bibitem[{{Kaltenegger} \& {Sasselov}(2011)}]{KalteneggerSasselov2011}
{Kaltenegger}, L., \& {Sasselov}, D. 2011, The Astrophysical Journal Letters,
  736, L25, \dodoi{10.1088/2041-8205/736/2/L25}

\bibitem[{{Kane} {et~al.}(2021){Kane}, {Arney}, {Byrne}, {Dalba}, {Desch},
  {Horner}, {Izenberg}, {Mandt}, {Meadows}, \& {Quick}}]{2021JGRE..12606643K}
{Kane}, S.~R., {Arney}, G.~N., {Byrne}, P.~K., {et~al.} 2021, J Geophys Res
  Planets, 126, e06643, \dodoi{10.1029/2020JE006643}

\bibitem[{Kasting(1988)}]{kasting-1988}
Kasting, J.~F. 1988, Icarus, 74, 472, \dodoi{10.1016/0019-1035(88)90116-9}

\bibitem[{Kasting {et~al.}(1993)Kasting, Whitmire, \&
  Reynolds}]{kasting1993habitable}
Kasting, J.~F., Whitmire, D.~P., \& Reynolds, R.~T. 1993, Icarus, 101, 108

\bibitem[{{Kempton} \& {Knutson}(2024)}]{2024RvMG...90..411K}
{Kempton}, E. M.~R., \& {Knutson}, H.~A. 2024, Reviews in Mineralogy and
  Geochemistry, 90, 411, \dodoi{10.2138/rmg.2024.90.12}

\bibitem[{{Kite} \& {Schaefer}(2021)}]{2021ApJ...909L..22K}
{Kite}, E.~S., \& {Schaefer}, L. 2021, \apjl, 909, L22,
  \dodoi{10.3847/2041-8213/abe7dc}

\bibitem[{{Koll} \& {Cronin}(2019)}]{2019ApJ...881..120K}
{Koll}, D. D.~B., \& {Cronin}, T.~W. 2019, \apj, 881, 120,
  \dodoi{10.3847/1538-4357/ab30c4}

\bibitem[{Komabayasi(1967)}]{komabayasi-1967}
Komabayasi, M. 1967, Journal of the Meteorological Society of Japan, 45, 137

\bibitem[{Kopparapu {et~al.}(2013)Kopparapu, Ramirez, Kasting, Eymet, Robinson,
  Mahadevan, Terrien, Domagal-Goldman, Meadows, \& Deshpande}]{Kopparapu_2013}
Kopparapu, R.~K., Ramirez, R., Kasting, J.~F., {et~al.} 2013, The Astrophysical
  Journal, 765, 131, \dodoi{10.1088/0004-637X/765/2/131}

\bibitem[{{Krijt} {et~al.}(2023){Krijt}, {Kama}, {McClure}, {Teske}, {Bergin},
  {Shorttle}, {Walsh}, \& {Raymond}}]{2023ASPC..534.1031K}
{Krijt}, S., {Kama}, M., {McClure}, M., {et~al.} 2023, in \ASPCS, Vol. 534,
  Astronomical Society of the Pacific Conference Series, 1031

\bibitem[{{Krissansen-Totton} {et~al.}(2024){Krissansen-Totton}, {Wogan},
  {Thompson}, \& {Fortney}}]{2024NatCo..15.8374K}
{Krissansen-Totton}, J., {Wogan}, N., {Thompson}, M., \& {Fortney}, J.~J. 2024,
  Nature Communications, 15, 8374, \dodoi{10.1038/s41467-024-52642-6}

\bibitem[{{Labrosse} {et~al.}(2007){Labrosse}, {Hernlund}, \&
  {Coltice}}]{2007Natur.450..866L}
{Labrosse}, S., {Hernlund}, J.~W., \& {Coltice}, N. 2007, \nat, 450, 866,
  \dodoi{10.1038/nature06355}

\bibitem[{Lambrechts {et~al.}(2019)Lambrechts, Morbidelli, Jacobson, Johansen,
  Bitsch, Izidoro, \& Raymond}]{Lambrechts2019b}
Lambrechts, M., Morbidelli, A., Jacobson, S.~A., {et~al.} 2019, \aap, 627, A83,
  \dodoi{10.1051/0004-6361/201834229}

\bibitem[{{Lebrun} {et~al.}(2013){Lebrun}, {Massol}, {Chassefi{\`e}Re},
  {Davaille}, {Marcq}, {Sarda}, {Leblanc}, \& {Brandeis}}]{2013JGRE..118.1155L}
{Lebrun}, T., {Massol}, H., {Chassefi{\`e}Re}, E., {et~al.} 2013, J Geophys Res
  Planets, 118, 1155, \dodoi{10.1002/jgre.20068}

\bibitem[{{L{\'e}ger} {et~al.}(2009){L{\'e}ger}, {Rouan}, {Schneider}, {Barge},
  {Fridlund}, {Samuel}, {Ollivier}, {Guenther}, {Deleuil}, {Deeg}, {Auvergne},
  {Alonso}, {Aigrain}, {Alapini}, {Almenara}, {Baglin}, {Barbieri}, {Bruntt},
  {Bord{\'e}}, {Bouchy}, {Cabrera}, {Catala}, {Carone}, {Carpano}, {Csizmadia},
  {Dvorak}, {Erikson}, {Ferraz-Mello}, {Foing}, {Fressin}, {Gandolfi},
  {Gillon}, {Gondoin}, {Grasset}, {Guillot}, {Hatzes}, {H{\'e}brard}, {Jorda},
  {Lammer}, {Llebaria}, {Loeillet}, {Mayor}, {Mazeh}, {Moutou}, {P{\"a}tzold},
  {Pont}, {Queloz}, {Rauer}, {Renner}, {Samadi}, {Shporer}, {Sotin}, {Tingley},
  {Wuchterl}, {Adda}, {Agogu}, {Appourchaux}, {Ballans}, {Baron}, {Beaufort},
  {Bellenger}, {Berlin}, {Bernardi}, {Blouin}, {Baudin}, {Bodin}, {Boisnard},
  {Boit}, {Bonneau}, {Borzeix}, {Briet}, {Buey}, {Butler}, {Cailleau},
  {Cautain}, {Chabaud}, {Chaintreuil}, {Chiavassa}, {Costes}, {Cuna Parrho},
  {de Oliveira Fialho}, {Decaudin}, {Defise}, {Djalal}, {Epstein}, {Exil},
  {Faur{\'e}}, {Fenouillet}, {Gaboriaud}, {Gallic}, {Gamet}, {Gavalda},
  {Grolleau}, {Gruneisen}, {Gueguen}, {Guis}, {Guivarc'h}, {Guterman},
  {Hallouard}, {Hasiba}, {Heuripeau}, {Huntzinger}, {Hustaix}, {Imad},
  {Imbert}, {Johlander}, {Jouret}, {Journoud}, {Karioty}, {Kerjean},
  {Lafaille}, {Lafond}, {Lam-Trong}, {Landiech}, {Lapeyrere}, {Larqu{\'e}},
  {Laudet}, {Lautier}, {Lecann}, {Lefevre}, {Leruyet}, {Levacher}, {Magnan},
  {Mazy}, {Mertens}, {Mesnager}, {Meunier}, {Michel}, {Monjoin}, {Naudet},
  {Nguyen-Kim}, {Orcesi}, {Ottacher}, {Perez}, {Peter}, {Plasson}, {Plesseria},
  {Pontet}, {Pradines}, {Quentin}, {Reynaud}, {Rolland}, {Rollenhagen},
  {Romagnan}, {Russ}, {Schmidt}, {Schwartz}, {Sebbag}, {Sedes}, {Smit},
  {Steller}, {Sunter}, {Surace}, {Tello}, {Tiph{\`e}ne}, {Toulouse}, {Ulmer},
  {Vandermarcq}, {Vergnault}, {Vuillemin}, \& {Zanatta}}]{2009A&A...506..287L}
{L{\'e}ger}, A., {Rouan}, D., {Schneider}, J., {et~al.} 2009, \aap, 506, 287,
  \dodoi{10.1051/0004-6361/200911933}

\bibitem[{{Li} {et~al.}(2015){Li}, {Dasgupta}, \& {Tsuno}}]{Li2015}
{Li}, Y., {Dasgupta}, R., \& {Tsuno}, . 2015, Earth and Planetary Science
  Letters, 415, 54, \dodoi{10.1016/j.epsl.2015.01.017}

\bibitem[{Libourel {et~al.}(2003)Libourel, Marty, \&
  Humbert}]{LIBOUREL20034123}
Libourel, G., Marty, B., \& Humbert, F. 2003, Geochimica et Cosmochimica Acta,
  67, 4123, \dodoi{https://doi.org/10.1016/S0016-7037(03)00259-X}

\bibitem[{{Lichtenberg}(2021)}]{2021ApJ...914L...4L}
{Lichtenberg}, T. 2021, \apjl, 914, L4, \dodoi{10.3847/2041-8213/ac0146}

\bibitem[{Lichtenberg {et~al.}(2021)Lichtenberg, Bower, Hammond, Boukrouche,
  Sanan, Tsai, \& Pierrehumbert}]{lichtenberg-2021}
Lichtenberg, T., Bower, D.~J., Hammond, M., {et~al.} 2021, Journal of
  Geophysical Research. Planets, 126, \dodoi{10.1029/2020je006711}

\bibitem[{{Lichtenberg} {et~al.}(2021){Lichtenberg}, {Bower}, {Hammond},
  {Boukrouche}, {Sanan}, {Tsai}, \& {Pierrehumbert}}]{lichtenberg_2021}
{Lichtenberg}, T., {Bower}, D.~J., {Hammond}, M., {et~al.} 2021, J Geophys Res
  Planets, 126, e06711, \dodoi{10.1029/2020JE006711}

\bibitem[{{Lichtenberg} \& {Clement}(2022)}]{2022ApJ...938L...3L}
{Lichtenberg}, T., \& {Clement}, M.~S. 2022, \apjl, 938, L3,
  \dodoi{10.3847/2041-8213/ac9521}

\bibitem[{{Lichtenberg} \& {Miguel}(2025)}]{2024arXiv240504057L}
{Lichtenberg}, T., \& {Miguel}, Y. 2025, Treatise on Geochemistry, 7, 51,
  \dodoi{10.1016/B978-0-323-99762-1.00122-4}

\bibitem[{{Lichtenberg} {et~al.}(2023){Lichtenberg}, {Schaefer}, {Nakajima}, \&
  {Fischer}}]{2023ASPC..534..907L}
{Lichtenberg}, T., {Schaefer}, L.~K., {Nakajima}, M., \& {Fischer}, R.~A. 2023,
  in Astronomical Society of the Pacific Conference Series, Vol. 534,
  Protostars and Planets VII, ed. S.~{Inutsuka}, Y.~{Aikawa}, T.~{Muto},
  K.~{Tomida}, \& M.~{Tamura}, 907, \dodoi{10.48550/arXiv.2203.10023}

\bibitem[{Marcq {et~al.}(2017)Marcq, Salvador, Massol, \&
  Davaille}]{marcq-2017}
Marcq, E., Salvador, A., Massol, H., \& Davaille, A. 2017, Journal of
  Geophysical Research. Planets, 122, 1539, \dodoi{10.1002/2016je005224}

\bibitem[{{Mlawer} {et~al.}(2023){Mlawer}, {Cady-Pereira}, {Mascio}, \&
  {Gordon}}]{2023JQSRT.30608645M}
{Mlawer}, E.~J., {Cady-Pereira}, K.~E., {Mascio}, J., \& {Gordon}, I.~E. 2023,
  \jqsrt, 306, 108645, \dodoi{10.1016/j.jqsrt.2023.108645}

\bibitem[{{Moore} {et~al.}(2023){Moore}, {Cowan}, \&
  {Boukar{\'e}}}]{2023MNRAS.526.6235M}
{Moore}, K., {Cowan}, N.~B., \& {Boukar{\'e}}, C.-{\'E}. 2023, \mnras, 526,
  6235, \dodoi{10.1093/mnras/stad3138}

\bibitem[{Morbidelli \& Nesvorny(2012)}]{Morbidelli2012a}
Morbidelli, A., \& Nesvorny, D. 2012, \aap, 546, 1,
  \dodoi{10.1051/0004-6361/201219824}

\bibitem[{Nakajima {et~al.}(2021)Nakajima, Golabek, Wünnemann, Rubie, Burger,
  Melosh, Jacobson, Manske, \& Hull}]{nakajima-2021}
Nakajima, M., Golabek, G.~J., Wünnemann, K., {et~al.} 2021, Earth and
  Planetary Science Letters, 568, 116983, \dodoi{10.1016/j.epsl.2021.116983}

\bibitem[{{Nakajima} {et~al.}(1992){Nakajima}, {Hayashi}, \&
  {Abe}}]{1992JAtS...49.2256N}
{Nakajima}, S., {Hayashi}, Y.-Y., \& {Abe}, Y. 1992, Journal of the Atmospheric
  Sciences, 49, 2256, \dodoi{10.1175/1520-0469(1992)049<2256:ASOTGE>2.0.CO;2}

\bibitem[{{Nicholls} {et~al.}(2024){Nicholls}, {Lichtenberg}, {Bower}, \&
  {Pierrehumbert}}]{harrison-2024}
{Nicholls}, H., {Lichtenberg}, T., {Bower}, D.~J., \& {Pierrehumbert}, R. 2024,
  Journal of Geophysical Research (Planets), 129, 2024JE008576,
  \dodoi{10.1029/2024JE008576}

\bibitem[{Nicholls {et~al.}(2024)Nicholls, Pierrehumbert, Lichtenberg,
  Soucasse, \& Smeets}]{nicholls_convective_2024}
Nicholls, H., Pierrehumbert, R.~T., Lichtenberg, T., Soucasse, L., \& Smeets,
  S. 2024, Monthly Notices of the Royal Astronomical Society, 536, 2957,
  \dodoi{10.1093/mnras/stae2772}

\bibitem[{{Olson} {et~al.}(2022){Olson}, {Sharp}, \&
  {Garai}}]{2022E&PSL.58717537O}
{Olson}, P., {Sharp}, Z., \& {Garai}, S. 2022, EPSL, 587, 117537,
  \dodoi{10.1016/j.epsl.2022.117537}

\bibitem[{{Pierrehumbert} \& {Gaidos}(2011)}]{2011ApJ...734L..13P}
{Pierrehumbert}, R., \& {Gaidos}, E. 2011, \apjl, 734, L13,
  \dodoi{10.1088/2041-8205/734/1/L13}

\bibitem[{Pierrehumbert(2010)}]{pierrehumbert-2010}
Pierrehumbert, R.~T. 2010, {Principles of Planetary Climate} (Cambridge
  University Press), \dodoi{10.1017/cbo9780511780783}

\bibitem[{{Pierrehumbert}(2010)}]{2010ppc..book.....P}
{Pierrehumbert}, R.~T. 2010, {Principles of Planetary Climate}

\bibitem[{{Quintana} {et~al.}(2016){Quintana}, {Barclay}, {Borucki}, {Rowe}, \&
  {Chambers}}]{2016ApJ...821..126Q}
{Quintana}, E.~V., {Barclay}, T., {Borucki}, W.~J., {Rowe}, J.~F., \&
  {Chambers}, J.~E. 2016, \apj, 821, 126, \dodoi{10.3847/0004-637X/821/2/126}

\bibitem[{{Ramirez}(2020)}]{2020MNRAS.494..259R}
{Ramirez}, R.~M. 2020, \mnras, 494, 259, \dodoi{10.1093/mnras/staa603}

\bibitem[{{Ramirez} \& {Kaltenegger}(2014)}]{2014ApJ...797L..25R}
{Ramirez}, R.~M., \& {Kaltenegger}, L. 2014, \apjl, 797, L25,
  \dodoi{10.1088/2041-8205/797/2/L25}

\bibitem[{{Ramirez} \& {Kaltenegger}(2017)}]{2017ApJ...837L...4R}
---. 2017, \apjl, 837, L4, \dodoi{10.3847/2041-8213/aa60c8}

\bibitem[{{Ramirez} \& {Kaltenegger}(2018)}]{2018ApJ...858...72R}
---. 2018, \apj, 858, 72, \dodoi{10.3847/1538-4357/aab8fa}

\bibitem[{Salvador {et~al.}(2017)Salvador, Massol, Davaille, Marcq, Sarda, \&
  Chassefi{\`e}re}]{salvador2017relative}
Salvador, A., Massol, H., Davaille, A., {et~al.} 2017, Journal of Geophysical
  Research: Planets, 122, 1458

\bibitem[{{Salvador} {et~al.}(2023){Salvador}, {Avice}, {Breuer}, {Gillmann},
  {Lammer}, {Marcq}, {Raymond}, {Sakuraba}, {Scherf}, \&
  {Way}}]{2023SSRv..219...51S}
{Salvador}, A., {Avice}, G., {Breuer}, D., {et~al.} 2023, \ssr, 219, 51,
  \dodoi{10.1007/s11214-023-00995-7}

\bibitem[{{Schaefer} \& {Elkins-Tanton}(2018)}]{2018RSPTA.37680109S}
{Schaefer}, L., \& {Elkins-Tanton}, L.~T. 2018, Philosophical Transactions of
  the Royal Society of London Series A, 376, 20180109,
  \dodoi{10.1098/rsta.2018.0109}

\bibitem[{{Schaefer} \& {Fegley}(2017)}]{2017ApJ...843..120S}
{Schaefer}, L., \& {Fegley}, Bruce, J. 2017, \apj, 843, 120,
  \dodoi{10.3847/1538-4357/aa784f}

\bibitem[{{Schaefer} {et~al.}(2016){Schaefer}, {Wordsworth}, {Berta-Thompson},
  \& {Sasselov}}]{2016ApJ...829...63S}
{Schaefer}, L., {Wordsworth}, R.~D., {Berta-Thompson}, Z., \& {Sasselov}, D.
  2016, \apj, 829, 63, \dodoi{10.3847/0004-637X/829/2/63}

\bibitem[{{Schlecker} {et~al.}(2024){Schlecker}, {Apai}, {Lichtenberg},
  {Bergsten}, {Salvador}, \& {Hardegree-Ullman}}]{2024PSJ.....5....3S}
{Schlecker}, M., {Apai}, D., {Lichtenberg}, T., {et~al.} 2024, Planet Sci, 5,
  3, \dodoi{10.3847/PSJ/acf57f}

\bibitem[{Schlecker {et~al.}(2021)Schlecker, Mordasini, Emsenhuber, Klahr,
  Henning, Burn, Alibert, \& Benz}]{Schlecker2021}
Schlecker, M., Mordasini, C., Emsenhuber, A., {et~al.} 2021, \aap, 656, A71,
  \dodoi{10.1051/0004-6361/202038554}

\bibitem[{{Schlichting} \& {Young}(2022)}]{2022PSJ.....3..127S}
{Schlichting}, H.~E., \& {Young}, E.~D. 2022, \psj, 3, 127,
  \dodoi{10.3847/PSJ/ac68e6}

\bibitem[{{Selsis} {et~al.}(2007){Selsis}, {Kasting}, {Levrard}, {Paillet},
  {Ribas}, \& {Delfosse}}]{selsis2007b}
{Selsis}, F., {Kasting}, J.~F., {Levrard}, B., {et~al.} 2007, Astronomy \&
  Astrophysics, 476, 1373, \dodoi{10.1051/0004-6361:20078091}

\bibitem[{Selsis {et~al.}(2023)Selsis, Leconte, Turbet, Chaverot, \&
  Bolmont}]{selsis-2023}
Selsis, F., Leconte, J., Turbet, M., Chaverot, G., \& Bolmont, E. 2023, Nature,
  620, 287, \dodoi{10.1038/s41586-023-06258-3}

\bibitem[{{Shorttle} {et~al.}(2024){Shorttle}, {Jordan}, {Nicholls},
  {Lichtenberg}, \& {Bower}}]{2024ApJ...962L...8S}
{Shorttle}, O., {Jordan}, S., {Nicholls}, H., {Lichtenberg}, T., \& {Bower},
  D.~J. 2024, \apjl, 962, L8, \dodoi{10.3847/2041-8213/ad206e}

\bibitem[{{Sossi} {et~al.}(2020){Sossi}, {Burnham}, {Badro}, {Lanzirotti},
  {Newville}, \& {O'Neill}}]{2020SciA....6.1387S}
{Sossi}, P.~A., {Burnham}, A.~D., {Badro}, J., {et~al.} 2020, Sci Adv, 6,
  eabd1387, \dodoi{10.1126/sciadv.abd1387}

\bibitem[{Sossi {et~al.}(2023)Sossi, Tollan, Badro, \& Bower}]{sossi-2023}
Sossi, P.~A., Tollan, P. M.~E., Badro, J., \& Bower, D.~J. 2023, Earth and
  Planetary Science Letters, 601, 117894, \dodoi{10.1016/j.epsl.2022.117894}

\bibitem[{{Suer} {et~al.}(2023){Suer}, {Jackson}, {Grewal}, {Dalou}, \&
  {Lichtenberg}}]{2023FrEaS..1159412S}
{Suer}, T.-A., {Jackson}, C., {Grewal}, D.~S., {Dalou}, C., \& {Lichtenberg},
  T. 2023, Front Earth Sci, 11, 1159412, \dodoi{10.3389/feart.2023.1159412}

\bibitem[{Tonks \& Melosh(1993)}]{tonks-1993}
Tonks, W.~B., \& Melosh, H.~J. 1993, Journal of Geophysical Research, 98, 5319,
  \dodoi{10.1029/92je02726}

\bibitem[{{Turbet} {et~al.}(2021){Turbet}, {Bolmont}, {Chaverot}, {Ehrenreich},
  {Leconte}, \& {Marcq}}]{2021Natur.598..276T}
{Turbet}, M., {Bolmont}, E., {Chaverot}, G., {et~al.} 2021, \nat, 598, 276,
  \dodoi{10.1038/s41586-021-03873-w}

\bibitem[{{Turbet} {et~al.}(2020){Turbet}, {Bolmont}, {Ehrenreich}, {Gratier},
  {Leconte}, {Selsis}, {Hara}, \& {Lovis}}]{2020A&A...638A..41T}
{Turbet}, M., {Bolmont}, E., {Ehrenreich}, D., {et~al.} 2020, \aap, 638, A41,
  \dodoi{10.1051/0004-6361/201937151}

\bibitem[{{Turbet} {et~al.}(2023){Turbet}, {Fauchez}, {Leconte}, {Bolmont},
  {Chaverot}, {Forget}, {Millour}, {Selsis}, {Charnay}, {Ducrot}, {Gillon},
  {Maurel}, \& {Villanueva}}]{2023A&A...679A.126T}
{Turbet}, M., {Fauchez}, T.~J., {Leconte}, J., {et~al.} 2023, \aap, 679, A126,
  \dodoi{10.1051/0004-6361/202347539}

\bibitem[{{Underwood} {et~al.}(2003){Underwood}, {Jones}, \&
  {Sleep}}]{UnderWoodJones2003}
{Underwood}, D.~R., {Jones}, B.~W., \& {Sleep}, P.~N. 2003, International
  Journal of Astrobiology, 2, 289, \dodoi{10.1017/S1473550404001715}

\bibitem[{{Walterov{\'a}} \&
  {B{\v{e}}hounkov{\'a}}(2020)}]{2020ApJ...900...24W}
{Walterov{\'a}}, M., \& {B{\v{e}}hounkov{\'a}}, M. 2020, \apj, 900, 24,
  \dodoi{10.3847/1538-4357/aba8a5}

\bibitem[{Wang {et~al.}(2018)Wang, Lineweaver, \& Ireland}]{WANG2018460}
Wang, H.~S., Lineweaver, C.~H., \& Ireland, T.~R. 2018, Icarus, 299, 460,
  \dodoi{https://doi.org/10.1016/j.icarus.2017.08.024}

\bibitem[{Warren(1985)}]{warren-1985}
Warren, P.~H. 1985, Annual review of earth and planetary sciences, 13, 201,
  \dodoi{10.1146/annurev.ea.13.050185.001221}

\bibitem[{{Wordsworth} \& {Kreidberg}(2022)}]{2022ARA&A..60..159W}
{Wordsworth}, R., \& {Kreidberg}, L. 2022, \araa, 60, 159,
  \dodoi{10.1146/annurev-astro-052920-125632}

\bibitem[{Wyatt {et~al.}(2019)Wyatt, Kral, \& Sinclair}]{Wyatt2019}
Wyatt, M.~C., Kral, Q., \& Sinclair, C.~A. 2019, Mon Not R Astron Soc, 802,
  782, \dodoi{10.1093/mnras/stz3052}

\bibitem[{{Yang} {et~al.}(2013){Yang}, {Cowan}, \&
  {Abbot}}]{2013ApJ...771L..45Y}
{Yang}, J., {Cowan}, N.~B., \& {Abbot}, D.~S. 2013, \apjl, 771, L45,
  \dodoi{10.1088/2041-8205/771/2/L45}

\bibitem[{{Zhang} \& {Yang}(2020)}]{zhangyang}
{Zhang}, Y., \& {Yang}, J. 2020, The Astrophysical Journal Letters, 901, L36,
  \dodoi{10.3847/2041-8213/abb87f}

\end{thebibliography}
\bibliographystyle{aasjournal}

\end{document}